\documentclass[a4paper,12pt]{article}

\usepackage{amssymb,latexsym,cite}

\makeatletter \@addtoreset{equation}{section} \makeatother

\begin{document}

\thispagestyle{empty}

\begin{flushright}
\hfill{HU-EP-04/29} \\
\hfill{MPP-2004-64}\\
\hfill{hep-th/0406118}
\end{flushright}

\vspace{15pt}

\begin{center}{\LARGE{\bf
Gaugino Condensation and 
Generation of \\[2mm]
Supersymmetric 3--Form Flux 
}}

\vspace{40pt}

{\bf G. L. Cardoso$^a$, G. Curio$^a$, G. Dall'Agata$^a$ and D.
L\"ust$^{a,b}$}

\vspace{10pt}

{\it  ${~}^a$ Humboldt Universit\"at zu Berlin,
Institut f\"ur Physik,\\
Newtonstrasse 15, D-12489 Berlin,
Germany}\\[1mm]
{E-mail: gcardoso,curio,dallagat,luest@physik.hu-berlin.de}

\vspace{10pt}

{\it  ${~}^b$ Max-Planck-Institut f\"ur Physik,\\
F\"ohringer Ring 6, D-80805 M\"unchen, Germany}\\[1mm]
{E-mail: luest@mppmu.mpg.de.de}

\vspace{50pt}

{ABSTRACT}

\end{center}

We extend the linearised solution of Polchinski and Strassler
describing the supergravity dual of the ${\cal N}=1^*$ gauge theory.
By analysing the equations of motion 
of type IIB supergravity at cubic order in the mass perturbation parameter,
we demonstrate the emergence of a 3--form flux of type $(3,0)$ with
respect to the natural complex structure. 
The generation of this flux can be associated to the 
dynamical formation of a gaugino condensate in the confining phase of
the ${\cal N}=1^*$ gauge theory.
We also check that the supersymmetry conditions are satisfied, and we
discuss how this  $(3,0)$--form flux is tied to the existence of
a supersymmetric background with $SU(2)$--structure.

\newpage

\baselineskip 6 mm

\section{Introduction}

The AdS/CFT correspondence \cite{Maldacena:1998re,Gubser:1998bc,Witten:1998qj} 
states that certain (supersymmetric) quantum field theories have a
dual description in terms of (super)gravity theories on backgrounds
which contain AdS factors.
An example thereof is provided by type IIB supergravity on an $AdS_5
\times S^5$ background with $N$ units of Ramond 5--form flux, which is
conjectured to be dual to ${\cal N}=4$ supersymmetric $SU(N)$ gauge
theory at large $N$ and at large 't Hooft coupling $g_{YM}^2N$.
This gauge theory, which is conformal, lives on a stack of $N$
D3--branes.
In subsequent studies, this conjecture has been generalised to a
correspondence between non--conformal gauge theories with less
supersymmetry and supergravity theories on deformed backgrounds.
An example of the latter is the type IIB supergravity solution of
Polchinski and Strassler \cite{Polchinski:2000uf}, which describes the
supergravity dual of the ${\cal N}=1^*$ $SU(N)$ gauge theory which is
obtained from the ${\cal N}=4$ $SU(N)$ gauge theory by mass
deformation, namely by giving mass $m$ to all three adjoint chiral
superfields.

\bigskip

The ${\cal
N}=1^*$ theory is interesting because it possesses confining phases
\cite{Vafa:1994tf,Donagi:1996cf,Strassler:1998ny,Dorey:1999sj,Dorey:2000fc}.
Moreover, for this theory 
the AdS/CFT correspondence allows for a quantitative
description of various non--perturbative field theory phenomena 
such as flux tubes, baryon vertices, domain walls,
instantons and condensates \cite{Polchinski:2000uf}.
In this paper, we will be concerned with the 
description of gaugino condensation.
As soon as the mass perturbation $m$ is turned on, a dynamical gaugino
condensate $< {\bar \lambda} {\bar \lambda}>\sim \Lambda^3$ gets
formed in the confining phase.  
Since, at large 't Hooft coupling $\Lambda = m$, the formation of a
gaugino condensate is an order $m^3$--effect in the dual supergravity
description.
According to the AdS/CFT correspondence, the formation of a gaugino
condensate (as well as its mass) corresponds, in the dual supergravity
solution, to the generation of a 2--form potential with polarisation
tensor $\varepsilon_{ijk}$ at order $m^3$ \cite{Polchinski:2000uf}.
It was shown in \cite{Polchinski:2000uf} that the linearised
equations of motion  do indeed allow for such a solution, albeit with an
undetermined coefficient
in front of it, whose precise value is 
in principle determined by the infrared physics.
On the other hand, when going beyond the linearised approximation and
thereby extending the analysis of  \cite{Polchinski:2000uf}, 
additional solutions are expected to arise at order $m^3$.
These solutions, being solutions to inhomogeneous equations, are uniquely
determined. 
It is in these inhomogeneous solutions that we will be interested in this paper.
They are the ones connected to the generation of 3--form flux in the
Polchinski--Strassler solution associated with the dynamical formation
of a gaugino condensate $<{\bar \lambda} {\bar \lambda}>$ in the
confining phase of the ${\cal N}=1^*$ gauge theory.

\bigskip

Inspired by analogy with compactifications of heterotic string theory
and of type IIB string theory on Calabi--Yau manifolds in the presence
of non--trivial fluxes \cite{Giddings:2001yu}, we will be looking for
the (3,0)--component of the 3--form flux $G_3$.
In heterotic string theory,
when considering a Minkowski ground state,
 the formation of a gaugino condensate
\cite{Ferrara:1983qs,Dine:1985rz,Derendinger:1985kk,Derendinger:1986cv,
Kounnas:1988ye,Font:1990nt,Ferrara:1990ei,Nilles:1990jv}
 requires the presence of a compensating 3--form H--flux of
Hodge types $(3,0)$ and $(0,3)$ (see also \cite{Cardoso:2003sp})
\footnote{
Additional supersymmetry preserving H--flux of Hodge types $(2,1)$ and
$(1,2)$ may be needed to compensate for non--K\"ahler deformations
compatible with an $SU(3)$--structure of the compactification manifold
\cite{Strominger:1986uh,Cardoso:2002hd,Becker:2003yv,Becker:2003gq,
Cardoso:2003af,Becker:2003sh,Gauntlett:2003cy}.}.  
On the other hand, in 
type IIB compactifications with D3--branes \cite{Camara:2003ku,Grana:2003ek},
it 
turns out that the effective soft SUSY breaking terms originating
from (3,0)--fluxes are of the type arising in dilaton dominated SUSY breaking
scenarios.  The latter also naturally arise in the context of spontaneous
supersymmetry breaking by gaugino condensation.
Actually, in \cite{Camara:2003ku,Grana:2003ek},
a direct link between the (3,0)--components of the
fluxes and the mass of the gaugini is established, and
this is suggestive of a similar relation for the gaugino condensate.
Of course, in 
heterotic string compactifications as well as in type IIB theory on compact
(warped) Calabi--Yau spaces, gravity cannot be decoupled as in the
AdS/CFT context, so that the resulting effective theories are {\sl locally}
supersymmetric field theories in which the formation of a gaugino
condensate, combined with the requirement of vanishing cosmological
constant, results in the spontaneous breaking of ${\cal N}=1 $
supersymmetry.  

\bigskip

In this paper, on the other hand, we will be dealing with a 
{\sl globally} supersymmetric field theory.
We will show that the formation of a 
gaugino condensate in such a field theory can be related,
through the AdS/CFT correspondence, to the emergence of 
a (3,0) $G_3$--flux in type IIB supergravity.
To be specific, we will show that when going beyond the linearised
analysis of the equations of motion given in \cite{Polchinski:2000uf},
a $(3,0)$--piece in the field strength $G_3$ {\sl does} get generated
to order $m^3$.
We will show that we may associate a 2--form potential with
polarisation tensor $\varepsilon_{ijk}$ to it, thereby associating the
emergence of this $(3,0)$--flux to the formation of a gaugino
condensate.
This $(3,0)$--form flux arises when taking into account that the
axion/dilaton field $\tau$ ceases to be constant at order $m^2$
\cite{Freedman:2000xb}, and solving the associated inhomogeneous
equations of motion for $G_3$.  
Note that at order $m^3$, the solution of the linearised equations of
motion alluded to above \cite{Polchinski:2000uf} does {\sl not} give
rise to a 3--form flux of Hodge type $(3,0)$.
Therefore, at order $m^3$, it is only when going beyond the linearised
approximation that we see the emergence of an imaginary anti--selfdual
3--form flux component $G^{(3,0)}$ of Hodge type $(3,0)$.

\bigskip

Since the ${\cal N}=1^*$ gauge theory is a {\sl globally}
supersymmetric field theory, the ground state described by the gaugino
condensate is ${\cal N}=1$ supersymmetric.
For this reason, and since the infrared vacuum is not
conformal, the associated 3--form flux on the dual supergravity side is
expected to be part of a supergravity background preserving ${\cal
N}=1$ supersymmetry.
This means that, in contrast to the (non--compact) 
Calabi--Yau solutions, the
emergence of $G^{(3,0)}$ should be compatible with an ${\cal N}=1$
residual supersymmetry of the background.
This may raise the issue of the consistency of such a background, since
supersymmetry is linked to the Hodge type of the 3--form flux and, 
for solutions preserving 4--dimensional Poincar\'e invariance,
the flux is usually of type (2,1) and/or (1,2).
This fact is tied to a specific spinor ansatz though, which is not
general enough to capture the above solution.
This ansatz reads
\begin{equation}
 \epsilon(x,y) = a(y) \, \varepsilon(x) \otimes \eta_-(y)
+ b(y) \, \varepsilon^*(x) \otimes \eta_+(y) \;,
\label{eq:typeC}
\end{equation}
where $a$ and $b$ denote complex functions, $\varepsilon$ is the
four--dimensional supersymmetry parameter and $\eta_{+}= (\eta_{-})^*$ 
is a globally defined spinor normalised to one.  
The existence of one globally defined spinor $\eta$ implies that the
tangent bundle over the transverse 6--dimensional space has an $SU(3)$
group structure.
The ansatz (\ref{eq:typeC}) includes the type A ansatz
\cite{Strominger:1986uh}, where $b = a^*$, the type B
\cite{Becker:1996gj,Grana:2000jj,Grana:2001xn}, where $b = 0$, and the
more general type discussed in \cite{Frey:2003sd}, which we call type
C.
In any of these cases, the allowed 3--form flux is constrained to contain only
(2,1) and/or (1,2) fluxes \cite{DallAgata:2004dk,Frey:2004rn}.
One may then ask how a 3--form flux of Hodge type $(3,0)$,
like the one which arises at order $m^3$ in the Polchinski--Strassler
background, can be compatible with ${\cal N}=1$ supersymmetry.
The solution to this question resides in using an even more general
supersymmetry ansatz, called type D \cite{DallAgata:2004dk}.
This ansatz reads
\begin{equation}
\quad \epsilon(x,y) = a(y) \, \varepsilon(x) \otimes
\eta_-(y) + \varepsilon^*(x) \otimes \left( b(y) \, \eta_+(y) + c(y) \,
\chi_+(y) \right)\;.
\label{typeD}
\end{equation}
It is based on the existence of two globally defined spinors, $\eta$
and $\chi$, which are linearly independent.
It implies that the group structure of the tangent bundle of the
transverse 6--dimensional space is further reduced to $SU(2)$.
Although this requirement is stronger than the one used to
formulate (\ref{eq:typeC}), it is necessary for obtaining more general
solutions, as it was already observed in type IIA
\cite{DallAgata:2003ir} and when considering $AdS_{5}$ solutions in
M--theory or type IIB \cite{Gauntlett:2004zh}.
In particular, using the ansatz (\ref{typeD}), the Hodge type of the
3--form flux is no more constrained by supersymmetry, and one can now
have (3,0) as well as (0,3) fluxes \cite{DallAgata:2004dk}.
Moreover, and in contrast with the case of $SU(3)$--structures, there is no 
preferred choice for the (almost) complex structure $J$, 
but one actually has a 
U(1)--worth of possibilities.
This would make the computation of the
Hodge type of the 3--form pointless.
However, for the solution at hand, there is a way to fix $J$, namely by the
fact that in the ultraviolet regime
we should recover the $AdS_5 \times S^5$ solution, 
which describes the near--horizon solution of a stack of D3--branes.
This means that in the ultraviolet the spinor ansatz should reduce to the
type B ansatz discussed above and therefore there is a unique
choice of $J$ associated to it.

\bigskip

There is an additional motivation for the fact that the ansatz D
(\ref{typeD}) is the appropriate one to describe the
Polchinski--Strassler solution and that the transverse 6--dimensional 
space displays an $SU(2)$--structure.
The addition of a mass perturbation to the chiral superfields
$\Phi_{i}$ implies that the F--term equations for a supersymmetric
vacuum read
$[\Phi_{i},\Phi_{j}] = - m\, \varepsilon_{ijk}\Phi_{k}$.
Since these fields have to be interpreted as the transverse
coordinates to the stack of D3--branes, this means that we may have
vacua where the D3--branes are spread over the transverse space building up 
a granular
$S^{2}$, i.e. they are polarised into 5--branes \cite{Myers:1999ps,Polchinski:2000uf}.
This obviously affects the spinor ansatz, and in addition to the usual 
projector on $\epsilon$ coming from the presence of the D3--branes one 
should have an additional one consistent with the 5--branes.
All in all, this means that one expects a supersymmetry projector of
the form \cite{Pilch:2004yg}
\begin{equation}
{\cal P} = \frac12 \left(1 - i
\gamma^{0}\gamma^{1}\gamma^{2}\gamma^{3}\left(\cos
\varphi + \sin \varphi \, \gamma^{4}\gamma^{5} *\right)\right)\,, \quad 
{\cal P} \epsilon = \epsilon,
\label{eq:projector}
\end{equation}
where $*$ denotes complex conjugation.
After some trivial algebra, it can be shown that this projector
naturally selects a spinor of the form given in (\ref{typeD}) with $b =
0$ (see also \cite{DallAgata:2004dk}).
It is therefore natural to expect that the full solution will possess
an $SU(2)$--structure.

\bigskip

Our findings may thus be summarised in the following  way,
\begin{center}
\centerline{$ m^3\; \longrightarrow \, G^{(3,0)} \;$  \qquad}
\centerline{\quad \quad  $\swarrow$ \quad $\searrow$
\phantom{$\displaystyle\int$}\quad}
\centerline{\fbox{$\; SU(2)$--structure } \quad \qquad
\fbox{ polarisation $\varepsilon_{ijk}\;$}\quad} 
\end{center}

The paper is organised as follows. In section 2 we review 
deformations of the AdS/CFT
correspondence and their relation to the harmonic analysis presented in
\cite{Kim:1985ez}.  In section 3 we establish the presence of a
$(3,0)$--piece in $G_3$ to order $m^3$, and we determine the
associated 2--form potential.  In section 4 we first compute the non--trivial
behaviour of the axion/dilaton $\tau$ to second order in $m$.  Then
we briefly review the impact of source terms on the Polchinski--Strassler
solution.  Next we turn to the explicit computation of various 
bulk contributions to $G_3$ to order $m^3$.  In section 5
we show that the emergence of a $(3,0)$-piece in $G_3$ is compatible
with supersymmetry based on the type D spinor ansatz.  We present
our conclusions in section 6.  Appendix A contains a summary of useful
expressions for the Polchinski--Strassler solution.

\section{Deformations of the AdS/CFT correspondence}

In this section we review the dictionary between field theory
operators and supergravity fields so that we can correctly identify
the couplings and vacuum expectation values (vevs) which are turned on
in our solution.

\bigskip

Consider an operator ${\cal O}$ of mass dimension $\Delta$.
It may be added to a four-dimensional CFT Hamiltonian,
\begin{equation} 
H = H_{\rm CFT} + a \, {\cal O} \;.
\label{deformo}
\end{equation}
The coupling $a$ has mass dimension $4- \Delta$.
The operator ${\cal O}$ may also develop a vev
\begin{equation} 
< 0 | {\cal O} | 0> = b \;. 
\label{vevo}
\end{equation} 
The vev $b$ has mass dimension $\Delta$.
In the dual supergravity description, $a$ and $b$ are read off
from the radial behaviour of the corresponding supergravity field at
large radius $r$,
\begin{equation} 
a \, \left(\frac{r}{R}\right)^{\Delta -4} + b \, R^{2 \Delta -4}
\, \left( \frac{r}{R}\right)^{- \Delta} \;. \label{abr} 
\end{equation}
Here $r^{\Delta -4}$ denotes the non-normalisable solution, whereas
$ r^{- \Delta}$ is the normalisable one.
$R= (4 \pi g N \alpha'^2)^{1/4}$
denotes the radius of $AdS_5$.  
Here, large $r$ corresponds to the ultraviolet regime in the dual
field theory.

\bigskip

The supergravity solution we will be considering is the type IIB
solution describing the supergravity dual of the ${\cal N}=1^*$ theory
of Polchinski and Strassler \cite{Polchinski:2000uf}.
The ${\cal N}=4$ supersymmetry of the unperturbed $SU(N)$ gauge theory
gets broken down to ${\cal N}=1^*$ by turning on mass terms for all
the three chiral superfields $\Phi_i$.
In the dual type IIB supergravity solution this corresponds to turning
on a 2--form potential with asymptotic behaviour (\ref{abr}), whose
field strength is related to the 3--form flux $G_3 = F_3 - \tau H_3$.
Here $F_3 = dC_2$ and $H_3 = d B_2$ denote the RR and the NSNS 3--form
field strengths, respectively, and $\tau = C + i {\rm e}^{-\Phi}$
denotes the type IIB axion/dilaton.
As argued in the introduction, also the gaugino condensate couples to 
the 2--form potential and this gives rise to an additional
contribution to the 3--form flux.
Let us then describe the correspondence for the 2--form potential in more
detail.

\bigskip

At large $r$, the supergravity solution goes back to $AdS_5 \times S^5$.
For this background the AdS/CFT correspondence has been explored in
great detail and all the possible field theory operators ${\cal
O}$ have been matched \cite{Witten:1998qj} with the Kaluza--Klein states appearing in the
analysis of \cite{Kim:1985ez}.
For our purpose, this implies that at least at the linearised level we
can read from the behaviour at large $r$ which couplings (\ref{deformo}) and
vevs (\ref{vevo}) get generated in the ${\cal N}=1^{*}$ theory.
The relation between the string quantities used here such as $G_3$ and 
the supergravity variables used in the KK reduction \cite{Kim:1985ez} 
of type IIB supergravity theory can be read off from\footnote{Note that (\ref{relpp}) differs
by a factor of $\sqrt{g}$ from the expression given in \cite{Grana:2001xn}.}
\begin{equation}
\kappa \, G_{SU} = i g^{1/2} \, \tau_2^{-1/2} \,
{\rm e}^{i \theta} \, G_3 \;\;\;,\;\;\;
{\rm e}^{i \theta} = \left( \frac{1 + i {\bar \tau}}{1 - i \tau}
\right)^{1/2} \;,
\label{relpp}
\end{equation}
where $G_{SU} = f ( d A_2 - B d {\bar A}_2)$ and $ \kappa \, A_2 = g^{1/2} 
(B_2 + i C_2)$.  
Here $B = (1 + i \tau)/(1 - i \tau)$ and $f^{-2} = 1 - |B|^2$.
For a constant axion/dilaton, i.e. $\tau = \tau_0
= C_0 + i g^{-1}$, so that $\tau_2 = g^{-1}$, 
we obtain the following relation between potentials 
\begin{equation}
f \, {\hat A}_2 - f B {\overline  {\hat A}}_2 = 
i \, {\rm e}^{i \theta} \, (C_2 - \tau_0 B_2) 
\;\;\;,\;\; {\hat A}_2 = B_2 + i C_2 \;.
\end{equation}
Setting
\begin{equation}
C_2 - \tau_0 B_2 = r^p \, S_2 \;,
\label{cbs}
\end{equation}
where $S_2$ denotes a 2--form, yields
\begin{equation}
f \, {\hat A}_2 - f B {\overline {\hat A}}_2 = i \, {\rm e}^{i \theta} \,
r^p \, S_2 \:.
\label{as}
\end{equation}
Inverting this relation we obtain
\begin{equation}
{\hat A}_2 = i \, r^p \, \left( {\rm e}^{i \theta} \, f \, S_2 - {\rm
e}^{-i \theta} \, f B \, {\bar S}_2 \right) \;,
\label{a2su}
\end{equation} 
which expresses the supergravity potential ${\hat A}_2$
in terms of $S_2 $ and ${\bar S}_2$.
Following \cite{Polchinski:2000uf}, the boundary operators we are
interested in couple to the 2--form potential $C_{2} - \tau_0 B_{2}$.
Nevertheless, we can use the
harmonic analysis given in \cite{Kim:1985ez} to identify them,
since this analysis was performed for $B = 0$,
which implies that the expansion of $\hat A_2$ and 
of $C_{2} - \tau_0 B_{2}$ is the same.
This is validated by the fact that the 2--form $r^p S_2$ appearing in 
(\ref{as}) is the lowest harmonic 2--form on $S^5$ and is given by
\begin{equation}
r^p S_2 
 = \frac12 \, r^{p+3} \, T_{mnp}\,  \left(\frac{x^m}{r} \right)
d\left(\frac{x^n}{r}\right) \wedge 
d \left(\frac{x^p}{r}\right) \;,
\label{a2}
\end{equation}
where $T_{mnp}$ denotes a constant 3--form tensor in 6 dimensions.
Comparison with (\ref{abr}) shows that this 2--form  
corresponds to an operator deformation when $p = \Delta
-7$, and to a vev of an operator when $p = - \Delta -3$.  

\bigskip

In the case of the Polchinski--Strassler solution, the 3--form $T_3 =
\frac16 \, T_{mnp} dx^m \wedge dx^n \wedge dx^p$ is taken to satisfy
$\star_6 T_3 = - i T_3$, which corresponds to the $\overline {\bf
10}$ representation of the $SO(6)$ tangent
space group \cite{Polchinski:2000uf}.  
The associated 2--form potential (\ref{cbs}) satisfies the linearised
bulk equation of motion for $G_3= d (C_2 - \tau_0 \,B_2)$,
\begin{equation}
d \left( Z^{-1} \, (\star_6 - i ) G_3 \right) = 0\;\;\;,\;\;\; Z =
\frac{R^4}{r^4} \;,
\label{homog}
\end{equation}
provided that $p=-4,-6$ \cite{Polchinski:2000uf}. 
The 2--form potential is thus given by (\ref{a2}) with $p =-4, -6$,
which describes the non-normalisable ($p=-4$) solution associated to
turning on an operator of dimension $\Delta = 3$, as well as the
normalisable ($p=-6$) solution associated with giving a vev to an
operator of dimension $\Delta = 3$.  
This 2--form potential is a harmonic 2--form on $S^5$ with
eigenvalue $M^2 = \Delta (\Delta -4)=-3$.
There are thus two homogeneous solutions to (\ref{homog}) given by $G_3
= 3 \, r^{-4} ( T_3 - \frac43 \, V_3)$ and $G_3 = 3\, r^{-6} (T_3 - 2
V_3)$, where $V_3 = d \log r \wedge S_2$ \cite{Polchinski:2000uf}.
In the following, we will write the latter as $G_3 = 3\, r^{-6} \left({\hat
T}_3 - 2 {\hat V}_3\right)$.
The constant tensor ${\hat T}_3$ can have different entries than
$T_3$.

\bigskip

Let us now see which are the corresponding operators.
Since $T_{3}$ is chosen in the $\overline {\bf 10}$ of $SU(4)$, the
corresponding operators should be in the ${\bf 10}$ and can be
identified with the fermionic bilinears of the ${\cal N} = 4$ 
theory, namely ${\cal O}^{10} = {\cal O}_{ij}^{6} + {\cal O}_{i}^{3} + {\cal 
O}^1 = \psi_i \psi_{j} + \lambda \psi_i + \lambda
\lambda$, where the splitting is done according to representations of the $SU(3)$ 
subgroup of $SU(4)$.
The ${\cal N}=1^*$ gauge theory is obtained by deforming the ${\cal
N}=4$ gauge theory by a mass deformation of the form $W = \frac12
m_{ij} \, \Phi_{(i} \Phi_{j)}$.
The superfield bilinear $\Phi_{(i} \Phi_{j)}$ (as well as ${\cal
O}_{ij}$) transforms as a $\bf 6$ under the $SU(3)$ subgroup of the
$SO(6)$ R--symmetry, whereas the mass matrix $m_{ij}$ transforms as a
$\bf \bar 6$, and it is this coupling which will be described by $T_3$.
The $\bf \bar 6$ can be represented by the primitive $(1,2)$--part
$T_3$ of an imaginary anti-selfdual 3--form, $\star_6 T_3 = - i T_3$.
Therefore, turning on equal mass terms for all the three chiral
superfields $\Phi_i$ corresponds, on the supergravity side, to turning
on the $p=-4$ solution $G_3 = 3\, r^{-4} ( T_3 - \frac43 \, V_3)$ with
$T_3$ being proportional to the mass parameter $m$
\cite{Polchinski:2000uf}.
Consequently, only some of the components of $T_{i \bar \jmath \bar k}$
are actually turned on.
The component $T_{ijk}$ is not turned on, since this would amount to
turning on a mass for the gaugini, thereby breaking supersymmetry.  

\bigskip

On the other hand, the $p=-6$ solution $G_3 =3\, r^{-6} ( {\hat T}_3 - 2
\, {\hat V}_3)$ may get turned on at order $m^3$, since ${\hat T}_3
\sim m^3$ on dimensional grounds.
As discussed above, it corresponds to turning on a vev of an operator
of dimension $\Delta = 3 $.
Note that the constant tensor ${\hat T}_3$ can have different entries
than the tensor ${T}_3$ appearing at linear order in $m$.
${\hat T}_3$ may, for instance, have a non-vanishing component ${\hat
T}_{ijk} \propto \varepsilon_{ijk}$.  
Since ${\hat T}_3$ is in the $\overline{\bf 10}$, the $SU(3)$--singlet
${\hat T}_{ijk}$ corresponds to a non-vanishing vev of the gaugino
condensate $\overline{\cal O} = {\bar \lambda } { \bar \lambda}$ in the confining phase of
the ${\cal N}=1^*$ theory. 
This is so, because the bilinear $\lambda \lambda$ belongs to the
${\bf 10}$, has mass dimension three and is a singlet under the
$SU(3)$--subgroup of the $SO(6)$ R--symmetry of the unperturbed ${\cal
N}=4$ theory.
Its vev is given by $\Lambda^3 = m^3 \, {\rm exp} ( - 8 \pi^2/
g^2_{\rm YM} N)$.  In the large 't Hooft-coupling limit we then have
$\Lambda^3 = m^3$, which is the behaviour of ${\hat T}_3$.
Note that the operator growing a vev is actually a linear combination
of ${\bar \lambda } { \bar \lambda}$, ${\bar \Phi}^3 $ and $m {\bar
\Phi}^2$, namely a primary operator orthogonal to the combination
appearing in the Konishi multiplet \cite{Ceresole:1999zs}.
By complex conjugation, also $\lambda \lambda$ grows a vev, which
corresponds to the entries of ${\overline{\hat T}_3}$.
The harmonic ${\bar S}_2$ in (\ref{a2su}) is therefore associated with
a vev of a linear combination of $\lambda \lambda$, ${\Phi}^3 $ and $m
{\Phi}^2$.
In the Higgs phase, on the other hand, there is no gaugino condensate
and hence no gaugino condensate vev.
However, a linear combination of ${\Phi}^3 $ and $m {\Phi}^2$ may have
a non-vanishing vev (these two vevs are in turn get related through
$< {\rm tr} \, \Phi \, \partial_{\Phi} W (\Phi)> =0$).
Therefore, also in the Higgs phase a homogeneous solution (\ref{a2})
with $p = -6$ may arise with a non-vanishing coefficient in front of
it.

\bigskip

Finally, observe that the solution $G_3 = 3\, r^{-6} \left({\hat T}_3 - 2
{\hat V}_3\right)$ satisfies $( \star_6 - i ) G_3 = 0$, so that it is in the
$(\star_6 + i)$-eigenspace.
Therefore, this solution does not contain any piece of Hodge type
$(3,0)$, even though ${\hat T}_3$ may have a non-vanishing singlet
component ${\hat T}_{ijk}$ (see (\ref{hodge}) for the action of
$(\star_6 \pm i)$ on the various Hodge types of a 3--form).  
In this paper we will show, however, that when going beyond the linearised
analysis of the equations of motion given in \cite{Polchinski:2000uf},
a $(3,0)$-component of $G_3$ gets generated at order $m^3$.  
This will be summarised in the next section.
This $(3,0)$--piece, which arises from the inhomogeneous solution of
the equation of motion for $G_{3}$, can be derived from a 2--form potential with
polarisation tensor $\varepsilon_{ijk}$, which is the polarisation
tensor associated with the gaugino condensate $<{\bar \lambda} {\bar
\lambda}>$ in the confining phase, thereby linking the emergence of a
$(3,0)$-piece in $G_3$ to the formation of a gaugino condensate.

\section{The $(3,0)$--part of the 3--form $G_3$}

At large $r$, the supergravity solution dual to the ${\cal N}=1^*$ theory 
admits an integrable complex structure in the transverse
6--dimensional space.
Locally, the associated fundamental 2--form $J$ may be written as $J =
e^1 \wedge e^2 + e^3 \wedge e^4 + e^5 \wedge e^6$, and $(1,0)$-forms
are given by $e^1 + i e^2$, $e^3 + i e^4$ and $e^5 + i e^6$.
Given a 3--form $G_3$, we may then project onto the various Hodge types of $G_3$ by
\begin{equation}
\begin{array}{rcl}
&&{(\star_6 - i) } \longrightarrow (3,0) + (1,2)_P + (2,1)_{NP} \;,
\\[2mm]
&&{(\star_6 + i) }\longrightarrow (0,3) + (2,1)_P+ (1,2)_{NP}
\;,\\[2mm]
\end{array}
\label{hodge} 
\end{equation} 
where the subscripts $P$ and $NP$ denote the primitive and non--primitive
parts, respectively.  
Here, $\star_6$ is defined in terms of $\epsilon_{abcdef}$ with $a,
\dots = 1, \dots, 6$.

\bigskip

To order $m$, the metric is given by
\begin{equation} ds^2_{10} = Z^{-1/2} \eta_{\mu \nu} dx^{\mu} dx^{ \nu} + Z^{1/2}
g_{mn} 
dx^m dx^n \;,
\label{metricm}
\end{equation}
where $g_{mn} = \delta_{mn}$.
We may introduce complex coordinates as
$e^1 + i e^2 = \sqrt{2} dz^1$, $e^3 + i e^4 = \sqrt{2} dz^2$ and
$e^5 + i e^6 = \sqrt{2} dz^3$, where \cite{Polchinski:2000uf}
\begin{equation} z^1 = \frac{x^4 + i
x^7}{\sqrt{2}} \;\;\;,\;\;\; z^2 = \frac{x^5 + i
  x^8}{\sqrt{2}} \;\;\;,\;\;\; z^3 = \frac{x^6 + i x^9}{\sqrt{2}} \;.
\label{zcoor}
\end{equation}
To order $m^3$, the metric $g_{mn}$ in the 6--dimensional transverse spaces
ceases to be diagonal \cite{Freedman:2000xb}, and therefore the
$(1,0)$--forms
will define new complex coordinates $v^1, v^2, v^3$. 
The differentials $dv^i$, when expressed in terms of 
$dz^i$ and $d{\bar z}^i$, may receive $d{\bar z}^i$--admixtures of order $m^2$.
Therefore, what for instance 
is a $(2,1)$--form when expressed in terms of the differentials
$dz^i$ and $d {\bar z}^i$, may become a $(3,0)$--form when expressed in terms
of the differentials $dv^i$.
In order to be sure that we correctly identify the $(3,0)$--contribution, 
we have to first understand if there are spurious
contributions to the $(3,0)$--part at order $m^{3}$ which come from the
order $m$ perturbation in $G_{3}$.

\bigskip

Since we will be working to order $m^3$, the $(3,0)$--part in question
will be the one given in terms of differentials $dv^i$.
We will, however, begin by using differentials $dz^i$ and $d {\bar
z}^i$.  
The 3--form flux $G_3 = G_{(1)} + G_{(3)}$ has order
$m$--contributions $G_{(1)}$ as well as order $m^3$--contributions
$G_{(3)}$.
The order $m$--contribution is given by the $p=-4$ solution discussed
in the previous section, $G_{(1)} = 3 r^{-4} \, ( T_3 - \frac43 V_3)$,
and it contains a primitive $(2,1)$-part $G^{(2,1)}_P$ (in
differentials $dz^i$ and $d {\bar z}^i$).  
This is the only contribution in $G_{(1)}$ which may turn into a
$(3,0)$--part (which we denote by $\Upsilon^{(3,0)}_v$) of order $m^3$
when expressed in terms of differentials $dv^i$.
We therefore write
\begin{equation}
G^{(2,1)}_P = \Upsilon^{(3,0)}_v + \dots \;.
\label{gg23}
\end{equation}
As we will be showing in this paper, the contribution $G_{(3)}$ is
given by the following expression,
\begin{equation}
G_{(3)} = Z \, U\, dz^1 \wedge dz^2 \wedge dz^3 - \frac{i}{2} ({\tilde
\star}_6 - \star_6) G^{(2,1)}_P + \dots \,,
\label{g3um}
\end{equation}
where $\star_6$ denotes the dual with respect to the flat metric
$\delta_{mn}$, whereas ${\tilde \star}_6$ denotes the dual with
respect to the curved (order $m^2$ corrected) metric $g_{mn}$.
Using (\ref{gg23}) and (\ref{hodge}), we obtain $\star_6 G^{(2,1)}_P =
i G^{(2,1)}_P = i \left(\Upsilon^{(3,0)}_v + \dots\right)$ as well as ${\tilde
\star}_6 G^{(2,1)}_P = {\tilde \star}_6 \left(\Upsilon^{(3,0)}_v + \dots\right)=
- i \Upsilon^{(3,0)}_v + \dots$, where the dots denote forms which are
not of the $(3,0)$-type in $dv^i$-differentials.
It follows that
\begin{equation}
G_{(3)} = Z \, U\, dv^1 \wedge dv^2 \wedge dv^3 - 
\Upsilon^{(3,0)}_v + \dots 
\label{g33}
\end{equation}
when expressed in terms of differentials $d v^i$.  
Adding up (\ref{gg23}) and (\ref{g33}), we find that the $(3,0)$--part
of $G_3$ (in differentials $dv^i$) is, to order $m^3$, given by
\begin{equation}
G^{(3,0)} = Z \, U\, dv^1 \wedge dv^2 \wedge dv^3 \;.
\label{30u}
\end{equation}
We find, moreover, that (\ref{30u}) is induced by the $(1,0)$--part of
$d \tau$ of order $m^2$.
This, as we will show, is in accordance with the supersymmetry
analysis given in \cite{DallAgata:2004dk}.

\bigskip

The $(3,0)$--piece (\ref{30u}) may be obtained from a 2--form potential
with polarisation tensor $\varepsilon_{ijk}$, as follows.  
The function $U$ is given by 
\begin{equation}
 U = \frac{1}{27 g}\, m^3\, R^4\, r^{-6} \, ({\bar z}^i)^2 \, ({\bar z}^j)^2\,,
\end{equation}
where $r^2 = 2 {\bar z}^i z^i$.
Note that to order $m^3$ we may simply use the coordinates $z^i$
instead of the $v^i$.  
Then, it can be checked that $G^{(3,0)}$ may be written as
\begin{equation}
G^{(3,0)} = \partial Y^{(2,0)}= - \frac{1}{108} \, g^{-1} \, m^3\, R^8
\, \partial \left( \frac{({\bar z}^i)^2 \, ({\bar z}^j)^2}{r^4} \, X_2
\right) \;,
\label{g30x2}
\end{equation}
where $X_2$ denotes the following 2--form potential
\begin{equation}
X_2 = r^{-6} \, \varepsilon_{i j k }
\, z^i dz^j \wedge dz^k \;,
\end{equation}
which satisfies $\partial X_2 = 0$.  
Observe that the 2--form potential $Y^{(2,0)}$ is not a harmonic
2--form on $S^5$, but this was not to be expected anyway since
$G^{(3,0)}$ does not arise as a solution to the linearised equations
of motion.

\bigskip

The homogeneous $p=-6$ solution given in (\ref{a2}) may also contain a
term with polarisation tensor $\varepsilon_{ijk}$, which when added to
$Y^{(2,0)}$, results in a potential\footnote{Note that in view of $dG_3 \neq
  0$, cf. (\ref{BIg3}), only a piece of $G_3$ can be related to a 2--form
potential.} given by
\begin{equation}
C_{2}-\tau_{0}B_{2}\, \sim \,m^3 \, r^{-6} \left( c + \frac{({\bar
z}^i)^2 \, ({\bar z}^j)^2}{r^4} \right) \varepsilon_{ijk} \, z^i \,
dz^j \wedge dz^k \;,
\end{equation}
where the constant $c$ denotes the contribution from the 
$p=-6$ solution.  
The constant $b$ appearing in (\ref{abr}) is then proportional to
$m^{3}\, c$.
Observe that the angular contribution $ ({\bar z}^i)^2 \, ({\bar
z}^j)^2 \, r^{-4}$ cannot be cancelled against the constant $c$ of the
homogeneous solution.
Since in the confining phase the polarisation tensor
$\varepsilon_{ijk}$ is associated to the vev $< {\bar \lambda} {\bar
\lambda}>$, we conclude that the $(3,0)$--part of $G_3$ contributes to
the vev of the gaugino condensate ${\bar \lambda} {\bar \lambda}$.
The
Higgs and confining phases are related by $g \leftrightarrow 1/g$
\cite{Polchinski:2000uf}.
In both phases there is the 2--form potential with polarisation tensor
$\varepsilon_{ijk}$.
In the Higgs phase this is related to the vev of ${\rm tr}\, \bar
\Phi^{3}$.

\bigskip

Actually, the above doesn't uniquely determine the potential
associated to $G^{(3,0)}$, since
$G^{(3,0)}$ may also be written as
\begin{equation}
G^{(3,0)} = - \frac{1}{432} \, g^{-1} \, m^3 \, R^8\, \partial \left(
\frac{({\bar z}^l)^2}{r^8} \, \varepsilon_{{\bar \imath} j k } \, {\bar
z}^i dz^j \wedge dz^k \right) \;,
\label{potmd}
\end{equation}
which results in a 2--form potential $P^{(2,0)}$ with a polarisation
tensor $\varepsilon_{{\bar \imath} jk}$.
The complex conjugate tensor $\varepsilon_{i {\bar \jmath} {\bar k} }$ is
proportional to $T_{i {\bar \jmath} {\bar k}}$ and therefore associated to
the mass deformation $W = \frac12 m_{ij} \, \Phi_{(i} \Phi_{j)}$ (note
that it cannot be associated to a vev of a bilinear in the matter
fermions, since such a vev would break supersymmetry).

\bigskip

Thus, the 2--form potential associated with $G^{(3,0)}$ is  
a linear combination of the potentials $Y^{(2,0)}$ and $P^{(2,0)}$.
In order to determine the precise linear combination, one would have
to fully determine $G_3$ at order $m^3$, which we haven't done, 
and to derive it from a 2--form
potential.
So, even though we cannot at this stage determine the precise linear
combination, the above shows that $G^{(3,0)}$ can be associated to the
vev of the gaugino condensate.

\section{Determining $G^{(3,0)}$}

\subsection{Bulk contribution to $\tau$}

It will turn out that the $(3,0)$--part of $G_3$ at order $m^3$
is driven by a non--constant axion/dilaton $\tau$ at order $m^2$.
In this section we will therefore determine the dependence
of $\tau $ on the six transverse coordinates $x^m$ at order $m^2$.  
The non--trivial behaviour of the dilaton
$\Phi (x^m)$ has already been determined in \cite{Freedman:2000xb}. 

\bigskip

To linear order in $m$, both the dilaton $\Phi$ and the axion $C$ are
constant \cite{Polchinski:2000uf}.
To quadratic order in $m$ this will not any longer be the case.

\bigskip

The bulk equation of motion for the dilaton reads \cite{Polchinski:2000uf} 
\begin{equation}
\nabla^M \nabla_M \Phi = {\rm e}^{2 \Phi} \, \partial_m C \partial^m C +
\frac{g {\rm e}^{\Phi}}{12} {\rm  Re} \,  ( G_{mnp} G^{mnp} ) \;.
\label{eqp} 
\end{equation} 
The bulk equation of motion for $C$ \cite{Polchinski:2000uf}, 
\begin{equation} 
\nabla^M ( {\rm e}^{2 \Phi} \partial_M C) = - \frac{g {\rm
e}^{\Phi}}{6} H_{mnp} {\tilde F}^{mnp} \;, \label{eqc} 
\end{equation}
shows that $\partial_m C$ is non-vanishing to second order in $m$,
i.e. $\partial_m C \sim m^2$.
Therefore $\partial_m C \partial^m C \sim m^4$, and we can neglect
this term in (\ref{eqp}) relative to $ G_{mnp} G^{mnp}$, which is of
order $m^2$.

\bigskip

Using the bulk expression for $G_3$ \cite{Polchinski:2000uf}, $G_3 = d
\eta_2 = g^{-1} \, Z \, (T_3 - \frac43 V_3 ) $, as well as
(\ref{metricm}) we compute \cite{Freedman:2000xb}
\begin{equation}
G_{mnp} \,  G^{mnp} = g^{-2} Z^{1/2} \left( T_{mnp} \,T_{mnp} -
\frac83 \frac{x^m x^n}{r^2} \, T_{mqp} \, T_{nqp} \right)\;, 
\end{equation}
where the contractions on the right hand side are with respect to the flat
metric $\delta_{mn}$.  
Using the definition of $T_3$ 
\begin{equation} 
T_3 = m \, \left( dz^1 \wedge d {\bar z}^2 \wedge d {\bar z}^3 + d
{\bar z}^1 \wedge d z^2 \wedge d {\bar z}^3 + d {\bar z}^1 \wedge d
{\bar z}^2 \wedge d z^3 \right) \;,
\label{t3}
\end{equation} 
which has Hodge type $(1,2)$ and is primitive, i.e. $T_3 \wedge J =
0$, we get
\begin{equation}
G_{mnp} \,  G^{mnp} = \frac{32 i}{3} \,
g^{-2} \, Z^{1/2} \, m^2 \, \left(\Sigma + i \frac{Y}{2} \right)\;.
\label{g2}
\end{equation}
Here $\Sigma$ and $Y$ denote two $SO(6)$ scalar harmonics satisfying
\begin{equation} 
\Delta_{\rm flat} H = - \frac{12}{r^2} \, H\;\;\;,\;\;\; H = \Sigma, Y
\;\;\;,
\end{equation} 
where $\Delta_{\rm flat}$ is the flat Laplacian in six dimensions.
$\Sigma $ is given by
\begin{equation} 
\Sigma \left(\frac{x^m}{r}\right) = \frac{x^4 x^7 + x^5 x^8 + x^6 x^9}{r^2} = -i
\frac{[(z^i)^2 -({\bar z}^i)^2] }{2r^2} \;,
\end{equation}
and $Y$ by
\begin{equation}
Y\left(\frac{x^m}{r}\right) = \frac{\sum_{i=4}^6 (x^i)^2 - \sum_{i=7}^9
(x^i)^2}{r^2} = \frac{(z^i)^2 +({\bar z}^i)^2 }{r^2} \;, 
\label{y} 
\end{equation}
where $r^2 = 2 z^i {\bar z}^i$.

\bigskip

The bulk equations of motion (\ref{eqp}) and (\ref{eqc})
can be solved at large $r$, where $Z = R^4/r^4$.
To order $m^2$, the bulk equation of motion (\ref{eqp}) becomes
\begin{equation}
Z^{-1/2} \,\Delta_{\rm flat} \Phi = \frac{g^2}{12} {\rm Re} \, (
G_{mnp} G^{mnp} ) \;,
\label{eqd} 
\end{equation} 
and it has been solved in \cite{Freedman:2000xb}. 
This equation has two homogeneous solutions $ \Phi = r^2 \,Y, r^{-6}
\,Y$, which are the non-normalisable and normalisable solutions for a
field operator of scale dimension $\Delta =6$ (see (\ref{abr})).
In the list of \cite{Kim:1985ez}, these homogeneous solutions
correspond to massive spin zero $k=2$ deformations in the $\bf 20_c$
representation of $SO(6)$ ($M^2 = k (k+4)= \Delta (\Delta -4)=12 $ in
the notation of \cite{Kim:1985ez}).
These homogeneous solutions will not be considered in the following.
\bigskip

Equation (\ref{eqd}) also has an inhomogeneous solution, which at large
$r$ reads\footnote{Note that (\ref{dilaz}) differs by a factor $18$
from the result given in \cite{Freedman:2000xb}.}
\cite{Freedman:2000xb}
\begin{equation}
\Phi - \Phi_0= \frac{m^2 R^4}{36 r^2} \, Y\left(\frac{x^m}{r}\right)
= \frac{m^2 R^4}{36 } \,
\frac{ \left[(z^i)^2 + ( {\bar z}^i)^2 \right]}{r^4} \;.
\label{dilaz} 
\end{equation} 
Note that $\Phi$ has mass dimension zero, as it should. 
It follows that \begin{equation} d
\Phi|_{(1,0)} = \frac{m^2 R^4}{18 r^4 } \, \left(
 z^j - 2\, \frac{{\bar z}^j}{r^2}
\, [(z^i)^2 + ( {\bar z}^i)^2 ] \right) \, dz^j \;. \label{ddil}
\end{equation}

The bulk equation of motion (\ref{eqc}) can, to order $m^2$, be
written as 
\begin{equation} 
Z^{-1/2} \,\Delta_{\rm flat} C = \frac{g}{6} \, {\rm Im} G_{mnp} \,
{\rm Re} G^{mnp} \;.  
\label{c2}
\end{equation} 
Using (\ref{g2}) and inserting the ansatz $C =  f(r) \,
\Sigma$ into (\ref{c2}) yields 
\begin{equation} 
\left( \frac{d^2}{d r^2} + \frac{5}{r} \frac{d}{dr} - \frac{12}{r^2}
\right) f (r) = \frac{8\, g^{-1} \, m^2 \, R^4}{9 \,r^4} \;.
\label{eqc2}
\end{equation} 
This has two homogeneous solutions of the form $f = r^2 , r^{-6}$,
which are the non-normalisable and normalisable solutions for a field
operator of scale dimension $\Delta =6$.
Again, these solutions will not be considered in the following.

\bigskip

Equation (\ref{eqc2}) also has an inhomogeneous solution given by
\begin{equation} 
f(r) = -\frac{g^{-1} \, m^2 \, R^4}{18 \,r^2} \;,
\end{equation} 
yielding
\begin{equation} 
g \, (C-C_0) =-\frac{ m^2 \, R^4}{18 \, r^2} \, \Sigma \;.
\end{equation}
Then, using $\tau = C + i {\rm e}^{- \Phi}$ and $\tau_0 = C_0+ i
g^{-1}$, we find \begin{equation} g \left(\tau - \tau_0\right)
=-\frac{ m^2 \, R^4}{18 \, r^2} \, \left( \Sigma + i \frac{Y}{2}
\right)= - i m^2\, R^4 \,\frac{({\bar z}^i)^2 }{18 \, r^4} \label{tau}
\end{equation} at large $r$.

\bigskip

The result (\ref{tau}) agrees with the analysis given in \cite{Grana:2000jj},
where supersymmetry was used to determine its general form.

\bigskip

Thus, to order $m^2$, the inhomogeneous solution $\tau - \tau_0$ is a
linear combination of the $SO(6)$ scalar harmonics $\Sigma$ and $Y$
with eigenvalue $M^2 = 12$.
This angular dependence is induced by $G^{mnp}G_{mnp}$.

\bigskip

In the following sections we will make use of the $(1,0)$--part of
(\ref{tau}), which reads
\begin{equation} 
g\, d \tau|_{(1,0)} = \frac{2i}{9} \, m^2 R^4 \frac{({\bar
z}^i)^2}{r^6} \, {\bar z}^j dz^j \;.
\label{dtau} 
\end{equation}

\subsection{Source terms}

We now discuss the contributions to $G_{3}$ due to the presence of sources
and argue that we can neglect them in the computation of $G^{(3,0)}$.

\bigskip

As shown in \cite{Polchinski:2000uf}, the Bianchi identity for $G_3$
gets modified by a magnetic source $J_4$, $d G_3 = J_4$, which is a
source for a D5/NS5--brane solution.  The latter is induced by the
Myers effect.
The source term $J_4$ is proportional to $ M \, \delta (w -r_0)$.
Here the coordinates $w^i$ are the imaginary part of the $z^i$, i.e.
$x^7, x^8, x^9$.
The quantity $M$ is given by $M = c\, \tau_0 + d$.  In the Higgs phase
$(c,d) = (0,1)$, so that $M=1$, whereas in the confining phase $(c,d)=
(1,0)$, so that $M = \tau_0$, where we may take $\tau_0 = i g^{-1}$.  
Therefore, in the Higgs phase, $J_4$ is real and hence $H_3=0$,
whereas in the confining phase $J_4$ is imaginary and hence $F_3=0$.
In the Higgs phase the D3--branes expand to a D5--brane with topology
$R^4 \times S^2$ , where the 2--sphere $S^2$ has radius $r_0 \sim m$.
The confining vacuum, on the other hand, is described by a NS5--brane
solution of the same topology and with radius $r_0 \sim g \,m$
\cite{Polchinski:2000uf}.  
Thus, $J_4$ has indeed the expected behaviour of a source for a
D5/NS5--brane solution.

\bigskip

The source modification of the Bianchi identity for $G_3$ results in a
modification of the sourceless order $m$ solution for $G_3$, namely
\cite{Polchinski:2000uf} $G_3 = d \eta_2 \rightarrow G_3 = d \eta_2 +
( \star_6 + i ) \, d \omega_2$.
Both $\eta_2$ and $\omega_2$ now depend on $r_0$.
Even though ultimately the two parameters $r_0$ and $m$ become linked,
as discussed above, it is helpful to think of $r_0$ and $m$ as
independent parameters in which one may power expand the solution.
Therefore, the solution constructed in \cite{Polchinski:2000uf} is
valid to linear order in $m$ and to any order in $r_0$.
This modification of $G_3$ is in agreement \cite{Grana:2000jj} with
supersymmetry, since it is of the form (\ref{grapol}), with $Z =
R^4/r^4$ replaced by $Z = R^4/(AB)$, where $A = y^2 + (w + r_0)^2$ and
$B = y^2 + (w - r_0)^2$, and where the coordinates $y^i$ denote
$x^4,x^5,x^6$.
Observe that at large $r$, $ d \eta_2$ goes as $r^{-4}$, whereas $ d
\omega_2$ behaves as $r^{-6}$.  
Therefore, asymptotically $d \omega_2$ is subleading relative to $d
\eta_2$.

\bigskip

Even though both $\eta_2$ and $\omega_2$ now depend on $r_0$, $Z^{-1}
(\star_6 -i) G_3$ is still constant and given by $Z^{-1} (\star_6 -i)
G_3 = Z^{-1} (\star_6 -i ) d \eta_2 = - \frac{2i}{3} g^{-1} T_3$
\cite{Polchinski:2000uf}.
Therefore, $(\star_6 -i) G_3$ is still of type $(1,2)$ and primitive.
Thus, to linear order in $m$ and to any order in $r_0$, there are no
terms in $G_3$ of Hodge type $(3,0)$.

\bigskip

A term in $G_3$ of type $(3,0)$ may arise from bulk effects.
In the following, we will show that bulk effects do give rise to a
$(3,0)$--term of order $m^3$.  
In doing so, we will be ignoring corrections due to sources, i.e. we
will be neglecting $r_0$--corrections.
This is consistent, since we will only be interested in the asymptotic
(large $r$) behaviour of the induced $(3,0)$-term in $G_3$.

\bigskip

Observe that when solving the equations of motion for $\tau$ to order
$m^2$, we also neglected the presence \cite{Polchinski:2000uf} of
source terms in the equation of motion for $\tau$.
This is again consistent in the large $r$ limit, where the corrections
due to the sources are subleading.

\subsection{Bulk contribution to $G_3$}

The 3--form $G_3$ has been determined in \cite{Polchinski:2000uf} to
linear order in $m$ and in the presence of a magnetic source $J_4$.
Here, we will determine some of the contributions to $G_3$ arising at
order $m^3$.
As explained previously, we will neglect $r_0$--contributions from the
magnetic source, which are subleading.

\bigskip

To order $m^3$, the bulk equation of motion for $G_3$ is not any longer
given by (\ref{homog}), but there are additional terms present which
are quadratic in the fluctuating fields, as follows.
{From} equation  (18) in \cite{Polchinski:2000uf} we can read off the bulk
equation of motion for $G_3 = F_3 - \tau H_3$,
\begin{equation} 
d \star G_3 = i g F_5 \wedge
G_3 - i {\rm e}^{\Phi} d \tau \wedge \star {\tilde F}_3 \;. 
\end{equation} 
Here and in what follows ${\tilde F}_{3} = F_{3} - C H_3$ and $\tilde 
F_5 = F_5- C_{2} \wedge H_3$.
Using $F_5 \wedge G_3 = {\tilde F}_5 \wedge G_3$ (valid whenever $G_3$
has only non-vanishing legs in the 6--dimensional transverse space),
we obtain
\begin{equation}
d \star G_3 = i g {\tilde F}_5 \wedge
G_3 - i {\rm e}^{\Phi} d \tau \wedge \star {\rm Re} \,G_3 \;.
\label{dsg} \end{equation} 
The five-form ${\tilde F}_5$ is given by ${\tilde F}_5 = (1 + \star)
\,d \chi_4 $, where $\chi_4 = g^{-1} Z^{-1} \,D(r) \, dx^0 \wedge dx^1
\wedge dx^2 \wedge dx^3 $.
Inspection of equation  (105) in \cite{Freedman:2000xb} shows that $D(r)$
receives a correction to order $m^2$.  At large $r$, $D(r)$ is given
by $D(r) = 1 + (7 m^2 R^4 r^{-2})/3$.
The explicit form of $D(r)$ will, however, not be needed in the
following.  
Thus we obtain
\begin{equation} 
g {\tilde F}_5 \wedge G_3 = d ( Z^{-1} D(r) ) \wedge G_3 \wedge dx^0
\wedge dx^1 \wedge dx^2 \wedge dx^3 \;.
\end{equation}
The ten--dimensional line element corrected by $m^{2}$ terms reads
\begin{equation} 
ds^2 = Z^{-\frac12}
\, A(r) \, \eta_{\mu \nu} \, dx^{\mu} dx^{\nu}  + Z^{\frac12} \,
g_{mn}\, dx^m dx^n\;.
\label{mome} 
\end{equation}
At order $m^2$, the metric $g_{mn}$ ceases to be diagonal, and
$A(r)$ receives a correction as well, which at large $r$ reads
$A(r) = 1 + (7 m^2 R^4 r^{-2})/24$ \cite{Freedman:2000xb}.
The explicit form of $A(r)$ will, however, not be needed in the
following.

\bigskip

{For} a metric of the form (\ref{mome}) we have
\begin{equation}
\star G_3 = Z^{-1} A^2 (r) ({\tilde \star}_6 G_3) \wedge dx^0 \wedge dx^1 
\wedge dx^2 \wedge dx^3 \;,
\end{equation} 
where ${\tilde \star}_6$ denotes the dual in the 6--dimensional
transverse space with respect to the corrected metric $g_{mn}$.  
The dual with respect to the flat 6--dimensional metric $\delta_{mn}$
will be denoted by $\star_6$.  
We then obtain from (\ref{dsg}) that 
\begin{equation} 
d \left( Z^{-1} A^2 (r) {\tilde \star}_6 G_3\right) = i d \left(
Z^{-1} D(r) \right) \wedge G_3 - i {\rm e}^{\Phi} d \tau \wedge
{\tilde \star}_6\left( Z^{-1} A^2 (r) \,{\rm Re} \,G_3 \right) \;.
\label{eqg2} 
\end{equation} 
In the presence of a non--trivial axion/dilaton field $\tau$, the bulk Bianchi
identity for $G_3$ reads
\begin{equation} 
d G_3 = - d\tau \wedge H_3 = - \frac{i}{2} \, g \, d \tau \wedge (G_3
- {\bar G}_3 ) \;.
\label{BIg3}
\end{equation} 
Thus we may rewrite (\ref{eqg2}) as 
\begin{eqnarray} 
d \left[ Z^{-1} A^2 (r) ({\tilde \star}_6 - i) G_3\right] &=& i d
\left[Z^{-1} (D(r) - A^2 (r) )\right] \wedge G_3+i Z^{-1} A^2 (r) d
\tau \wedge H_3\nonumber\\
&& - i {\rm e}^{\Phi} d \tau \wedge {\tilde \star}_6\left( Z^{-1}
A^2(r) \,{\rm Re} \,G_3 \right) \nonumber\\
&=&i d \left[Z^{-1} (D(r) - A^2 (r) )\right] \wedge G_3
 -{\rm e}^{\Phi} Z^{-1} A^2 (r) d \tau \wedge ({\tilde \star}_6 + i)
{\rm Im } \,G_3 \nonumber\\
&& - i {\rm e}^{\Phi} Z^{-1} A^2 (r) d \tau \wedge {\tilde \star}_6
\,G_3  \;. 
\label{eqg3a} 
\end{eqnarray}
We note that both $D - A^2$ and $d \tau$ are of order $m^2$.  We solve
(\ref{eqg3a}) to order $m^3$ by inserting the order $m$ value for
$G_3$ on the rhs of (\ref{eqg3a}).
To order $m^3$ (\ref{eqg3a}) becomes
\begin{eqnarray} 
d \left[ Z^{-1} A^2 (r) ({\tilde \star}_6 - i) G_3\right]
&=&i d \left[Z^{-1} (D(r) - A^2 (r) )\right] \wedge G_3 \nonumber\\
&&- g \,   Z^{-1} \, d \tau \wedge \left[(\star_6 + i)   {\rm Im } \,G_3 +
i \star_6 \,G_3 \right] \;. 
\label{eqg3} 
\end{eqnarray}
The order $m$ value for $G_3$ reads \cite{Polchinski:2000uf} $G_3=
(\star_6 + i) d \omega_2 + d \eta_2$.
The $\omega_2$-contribution stems from the inclusion of a magnetic
source in the Bianchi identity for $G_3$.
At large $r$, $d \omega_2$ is subleading relative to $d \eta_2$.
We therefore only keep the $d\eta_2$-terms in $G_3$, 
\begin{equation} 
G_3 \approx d \eta_2 = g^{-1} \, Z \, \left(T_3 - \frac43 V_3 \right)
\;.
\label{fps}
\end{equation}
In appendix \ref{g3m} we have listed various properties of $T_3$ and $V_3$.
Both $T_3$ and $V_3$ are written in terms of differentials $dz^i$ and
$d {\bar z}^i$.  
Observe that to order $m$, there are no $(3,0)$ and $(2,1)_{NP}$ parts
in (\ref{fps}).

\bigskip

Using $dr \wedge V_3 =0$ and (\ref{rela}), we obtain from (\ref{eqg3})
\begin{eqnarray} 
d \left[ Z^{-1} A^2 (r)\left( (\star_6 -i )\, G_3 + ({\tilde \star}_6
- \star_6) G_3 \right)\right] &=& i \,g^{-1} Z\, d \left[Z^{-1} (D(r)
- A^2 (r) )\right] \wedge T_3 \, \nonumber\\
&&- \frac13   \, d \tau \wedge \left(T_3 - {\bar T}_3 \right)\;.
\label{gggg} 
\end{eqnarray} 
Observe that $({\tilde \star}_6 - \star_6)$ is of order $m^2$.

\bigskip

We will now determine the $(3,0)$-part of the flux $G_3$
to order $m^3$.  
To do so, we will be interested in the $(3,1)$-part of equation
(\ref{gggg}).
On the right hand side, only the term proportional to $d \tau \wedge
{\bar T}_3$ contributes.
On the left hand side, the following terms may contribute.  {From}
(\ref{hodge}) we establish $(\star_6 -i )\, G_3 = - 2i \, (G^{(3,0)} +
G^{(2,1)}_{NP} + G^{(1,2)}_P)$.
To order $m$, the only non-vanishing piece in $(\star_6 -i )\, G_3$ is
the $G^{(1,2)}_P$-part.
Since $(A^2 -1)\, G^{(1,2)}_P$ is of type $(1,2)$, it will not
contribute to the $(3,1)$-part of equation (\ref{gggg}).
However, to order $m^3$, new pieces $G_{(3)}^{(3,0)}$ and $G_{(3)
NP}^{(2,1)}$ may get generated, and they will contribute to the
$(3,1)$-part of (\ref{gggg}) (a new $G_{(3) P}^{(1,2)}$ may also be
generated, but if we assume that the 6--dimensional transverse space
is complex, then this will not contribute to the $(3,1)$-part of
(\ref{gggg})).
Now consider the contribution $({\tilde \star}_6 - \star_6) \,G_3$.
To order $m$, the $(2,1)_{P}$-part of $G_3$ is non-vanishing, whereas
the $(2,1)_{NP}$-part vanishes.  Therefore, the only term in $({\tilde
\star}_6 - \star_6) \,G_3$ which may contain a $(3,0)$-piece to order
$m^3$ is the term $({\tilde \star}_6 - \star_6) G_P^{(2,1)}$.
To order $m$, there is also a non-vanishing $(1,2)$-part $G^{(1,2)}$.
Therefore, $({\tilde \star}_6 - \star_6) (G_P^{(2,1)} + G^{(1,2)})$
may contribute $(2,1)$-parts to order $m^3$.
Hence we infer from (\ref{gggg}) that to order $m^3$
\begin{equation} 
- 2i \,d \left.\left[ Z^{-1}\left( G_{(3)}^{(3,0)} + G^{(2,1)}_{(3)
NP} + \frac{i}{2} \left({\tilde \star}_6 - \star_6\right)\left(
G_P^{(2,1)} + G^{(1,2)}\right) \right)\right]\right|_{(3,1)} = \frac13
\, \left.d \tau \right|_{(1,0)} \wedge {\bar T}_3 \,.
\label{diffg30}
\end{equation}

Equation (\ref{diffg30}) may be solved as follows. 
 Using $d = \partial + {\bar \partial}$, we write ${\bar T}_3 = {\bar
 \partial} {\bar S_2}^{(2,0)} $, where
\begin{equation}
{\bar S_2}^{(2,0)} = \frac{m}{2} \, \varepsilon_{{\bar i}jk} \, {\bar
z}^i \, dz^j \wedge dz^k.
\label{eq:S2}
\end{equation}
Then, assuming that the 6--dimensional transverse space
is complex, we obtain
\begin{equation}
\left.d \tau \right|_{(1,0)} \wedge {\bar T}_3 = - {\bar \partial}
\left(\partial \tau \wedge {\bar S_2}^{(2,0)}\right) - \partial
\left({\bar \partial} \tau \wedge {\bar S_2}^{(2,0)} \right) \;.
\label{ts}
\end{equation} 
Introducing $U$ as 
\begin{equation}
G^{(3,0)}_{(3)} + \frac{i}{2} \left.\left(\left({\tilde \star}_6 - \star_6\right) G_P^{(2,1)}
\right)\right|_{(3,0)} = Z \, U \, dz^1 \wedge dz^2 \wedge dz^3 \; \,
\label{ansu}
\end{equation}
and $V = V_i\, dz^i $ as
\begin{equation}
G^{(2,1)}_{(3) NP} + \frac{i}{2}\left. \left(\left({\tilde \star}_6 -
\star_6\right) \left(G_P^{(2,1)}+ G^{(1,2)}\right)
\right)\right|_{(2,1)_{NP}} =- i \, Z \, V \wedge J
\;,
\label{ansv}
\end{equation} 
the lhs of (\ref{diffg30}) can be written as
\begin{equation}
- 2i {\bar \partial} \left( U dz^1 \wedge dz^2 \wedge dz^3\right) -2
\partial \left(V \wedge J \right) + \partial \left[ Z^{-1}
\left.\left(\left({\tilde \star}_6 - \star_6\right) \left(G_P^{(2,1)} + G^{(1,2)}\right)
\right)\right|_{(2,1)_{P}}\right] .
\label{uvp}
\end{equation}
Then, comparing (\ref{uvp}) with (\ref{ts}) yields
\begin{equation}
U dz^1 \wedge dz^2 \wedge dz^3 = - \frac{i}{6} \, \partial \tau 
\wedge {\bar S_2}^{(2,0)} \;,
\label{r1}
\end{equation}
as well as 
\begin{equation}
V \wedge J = \frac16 \left.\left({\bar \partial} \tau \wedge {\bar
S_2}^{(2,0)} \right)\right|_{NP} \;,
\label{r2}
\end{equation}
and also
\begin{equation}
Z^{-1} \, \left.\left(\left({\tilde \star}_6 - \star_6\right) \left(G_P^{(2,1)}
+ G^{(1,2)}\right)\right)
\right|_{(2,1)_{P}}
= - \frac13 \,
\left.\left({\bar \partial} \tau \wedge {\bar S_2}^{(2,0)}
\right)\right|_{P} \;,
\label{r3}
\end{equation}
up to various integration functions which we have set to zero.
The subscripts $NP$ and $P$ denote the non-primitive and primitive
parts, respectively.  

\bigskip

Observe that (\ref{ansu}) is of the form (\ref{g3um}).
It then follows from (\ref{30u}) that $Z U$ is the total induced
$(3,0)$-form at order $m^3$.
We therefore have
\begin{equation}
G^{(3,0)} =  Z U \,dz^1 \wedge dz^2 \wedge dz^3 = - \frac{i}{6} \, Z \,
{\bar S}_2^{(2,0)} \wedge d \tau|_{(1,0)} \;.
\label{g30ddil}
\end{equation}
Inserting (\ref{dtau}) into (\ref{g30ddil}) yields
\begin{equation} 
U = \frac{1}{27} \,g^{-1} m^3 R^4 \, \frac{ ({\bar z}^j)^2 \, ({\bar
z}^l)^2}{ r^{6}} \;.
\label{U}
\end{equation} 
In the next section we will show that $G^{(3,0)}$ is in precise
agreement with supersymmetry.

\bigskip

Using
\begin{equation}
\left.\left({\bar \partial} \tau \wedge {\bar S_2}^{(2,0)}
\right)\right|_{NP} = \frac{ g^{-1} m^3 \, R^4 }{9 } \, \frac{({\bar
z}^i)^2}{r^6} \, \varepsilon_{s{\bar t}k}\, z^s {\bar z}^t \, dz^k
\wedge J \;,
\end{equation}
we infer from (\ref{r2}) that
\begin{equation} 
V = \frac{ g^{-1} m^3 \, R^4 }{54 } \, \frac{({\bar z}^i)^2}{r^6} \,
\varepsilon_{s{\bar t}k}\, z^s {\bar z}^t \, dz^k\;.
\label{V}
\end{equation}

And finally, we note that (\ref{r3}) must be satisfied for consistency.
Although we did not perform this check explicitly, the alternative
derivation of (\ref{r1}) using supersymmetry enforces (\ref{r3}).

\bigskip

To summarise, we have found that to order $m^3$, there is a
$(3,0)$-piece in the flux $G_3$,
\begin{equation}
G^{(3,0)} = Z \, U\, dv^1 \wedge dv^2 \wedge dv^3 
= -\frac{g^{-1} m^3 \, R^8}{27 \, r^6} \, \left(\Sigma + i
\frac{Y}{2}\right)^2 \, dv^1 \wedge dv^2 \wedge dv^3 \;,
\label{g30u}
\end{equation}
which is induced by the $(1,0)$-part of $d \tau$ of order $m^2$.
This is a bulk effect.  
The fact that $G^{(3,0)}$ is generated by the $(1,0)$-part of $d \tau$
is in accordance with the supersymmetry analysis based on the type
D ansatz \cite{DallAgata:2004dk}, to which we now turn.

\section{Supersymmetry and $SU(2)$--structure}

We now check that the $(3,0)$-part of $G_3$ given in (\ref{g30ddil})
is consistent with supersymmetry.
The general analysis for the supersymmetry conditions in type IIB
based on the type D ansatz (\ref{typeD}) was performed in \cite{DallAgata:2004dk}.
This type D ansatz yields various restrictions on the allowed 3--form
flux.
Here we will check those conditions which are related to the
generation of $G^{(3,0)}$ to order $m^3$.
A check of the consistency of the Polchinski--Strassler background with
supersymmetry to order $m$ has been given in \cite{Grana:2000jj}.

\bigskip

We will be mainly concerned with the dilatino equations.
In particular, one of the supersymmetry conditions derived from this
in \cite{DallAgata:2004dk} states that the $(3,0)$-part of the flux
$G_3$ is entirely generated by part of the $(1,0)$-form $\partial
\tau$.  
The precise relation (for the signature used here) is as follows
\cite{DallAgata:2004dk}
\begin{equation}
\kappa \, g^{SU}_{3,0} = Z^{1/2}\, \frac{{\bar c}}{a} \, p_1 \;,
\label{gsug}
\end{equation}
where $ \kappa \, g^{SU}_{3,0} = i g \, {\rm e}^{i \theta} g_{3,0}$
and $G^{3,0} = g_{3,0} \, \Omega$, with $\Omega = 2 \sqrt{2}\, dz^1
\wedge dz^2 \wedge dz^3$.
The coefficients $a$ and $c$ are two of the coefficients appearing in
the type D supersymmetry spinor ansatz (\ref{typeD}).
In this ansatz, the two spinors $\eta_-$ and $\chi_+$ are related
by\footnote{The factor of $Z^{1/4}$ is due to the fact that here we
use $\gamma$--matrices satisfying (\ref{convg}).}
$\chi_+ = \frac12 Z^{1/4} \, w_i \, \gamma^i \, \eta_-$.
Here $w= w_m dx^m = w_i \, dz^i$ denotes a globally defined 1--form
necessary for the existence of an $SU(2)$--structure.
The coefficient $p_1$ is the one appearing in the decomposition 
\begin{equation}
P = P_m dx^m = p_1 w_m dx^m + p_2 {\bar w}_m dx^m + \Pi_m dx^m \;,
\label{P}
\end{equation}
where $P_m = f^2 \, \partial_m B = - {\rm e}^{2i \theta} \, \partial_m
\tau /(\tau - {\bar \tau})$ \cite{Grana:2001xn}.
The one-form $\Pi$ obeys $w \lrcorner\Pi = {\bar w} \lrcorner \Pi =0$.
Here we use the conventions of \cite{Grana:2000jj}, where $ds^2 = 
Z^{-1/2} \eta_{\mu \nu} dx^{\mu} dx^{\nu} + g_{mn} dx^{m} dx^n$ and
\begin{equation}
g_{i \bar j} = g_{{\bar j} i} = Z^{1/2} \, \delta_{i \bar j} \;\;\;,\;\;\;
\{\gamma^{i}, \gamma^{\bar j}\} = 2 g^{i \bar j} \;\;\;,\;\;\;
G^j\,_{jk} = g^{j {\bar i} } \, G_{{\bar i} jk} \;.
\label{convg}
\end{equation}
In these conventions, $w_i$ satisfies $w_i \, {\bar w}_{\bar i} \,
\delta^{i \bar i} = 2$.

\bigskip

To linear order in $m$, the coefficients $a,b$ and $c$ appearing in
(\ref{typeD}) can be determined from the result for the supersymmetry
spinor given in \cite{Grana:2000jj},
\begin{equation}
\epsilon (x^{\mu}, x^m) = Z^{-1/8} \, \varepsilon (x^{\mu})\otimes \eta_- 
+ \epsilon_1(x^{\mu}, x^m)\;,
\label{epsgp}
\end{equation}
where
\begin{equation}
\epsilon_1(x^{\mu}, x^m) = Z^{-1/8} \, \frac{\kappa}{24 S^2} \, (\partial_m 
\log Z )\, 
\varepsilon^* (x^{\mu})\otimes \gamma^m \, G_{SU}\, \eta_+  \;.
\label{eps1}
\end{equation}
Here $Z = R^4/r^4$ and $S^2 = 16\, Z^{- 1/2}\, r^{-2}$.  As before,
$\kappa \, G_{SU} = i g \, {\rm e}^{i \theta} \, G$, where $G= G_{mnp}
\, \gamma^{mnp}$.

\bigskip

{From} (\ref{epsgp}) we infer that $a = Z^{-1/8}$.  
In the following, we will show that $b=0$ and compute $c$, which we
will turn out to be non--vanishing.
This fact implies that the transverse 6--dimensional space possesses
an $SU(2)$--structure \cite{DallAgata:2004dk}.
We also note that since $c$ becomes non--trivial at order
$m$, in the $m \to 0$ limit we recover the type $B$ spinor ansatz as expected.
Moreover, the fact that $b = 0$ is compatible with the 
expectation that the dielectric configuration which underlies the
supergravity solution requires a supersymmetry spinor satisfying the
projector condition (\ref{eq:projector}).

\bigskip

Using
\begin{eqnarray}
\gamma_{ijk} \, \eta_+ &=& Z^{3/4} \, \Omega_{ijk} \, \eta_- = Z^{3/4}
\,2 \sqrt{2} \,\varepsilon_{ijk}\, \eta_- \;,\nonumber\\
\gamma^{{\bar i} {\bar k}} \, \eta_+ &=& \frac12 \, Z^{-1/4} \,
\delta^{{\bar i} i}\, \,\delta^{{\bar k} k}\, \Omega_{{ i} {k} {l}} \,
\gamma^{l} \, \eta_-
\end{eqnarray}
as well as
\begin{equation} 
G \, \eta_+  = G_{{\bar i}{\bar j}{\bar k}}\,
 \gamma^{{\bar i}{\bar j}{\bar k}} \,
\eta_+ + 6 G^{\bar j}\,_{{\bar j}{\bar k}} \, \gamma^{\bar k} \, \eta_+
\end{equation}
we obtain
\begin{eqnarray} 
\partial_m Z \, \gamma^m \, G \, \eta_+  &=&
\partial_l Z \, Z^{-3/4} \, G_{{\bar i}{\bar j}{\bar k}}\,
 \Omega_{ijk} \, \gamma^l  \eta_-
+ 12 \partial_i Z \, 
G^{\bar j}\,_{{\bar j}{\bar k}} \, g^{i {\bar k}} \, \eta_+
\nonumber\\
&&+
6 \partial_{\bar i} Z \, 
G^{\bar j}\,_{{\bar j}{\bar k}} \,  \gamma^{{\bar i} {\bar k}} \, \eta_+ 
\nonumber\\
&=& 2 \sqrt{2}\,
\partial_l Z \, Z^{-3/4} \, G_{{\bar i}{\bar j}{\bar k}}\,
 \varepsilon_{ijk} \, \gamma^l  \eta_-
+ 12 \partial_i Z \, 
G^{\bar j}\,_{{\bar j}{\bar k}} \, g^{i {\bar k}} \, \eta_+
\nonumber\\
&&+
6 \sqrt{2} \, \partial_{\bar i} Z \, Z^{-1/4} \, 
G^{\bar j}\,_{{\bar j}{\bar k}} \,  
\delta^{{\bar i} i}\,
\,\delta^{{\bar k} k}\,
\varepsilon_{{ i}
{k} {l}} \, \gamma^{l} \, \eta_- 
\;.
\end{eqnarray}
Using (\ref{vv}) we find that $ \partial_i Z \, G^{\bar j}\,_{{\bar
j}{\bar k}} \, g^{i {\bar k}} \, \eta_+ \propto {\bar z}^i\,
\varepsilon_{stk} \, z^s {\bar z}^t \, \delta^{i \bar k} =0$.  
It follows that
\begin{eqnarray}
\partial_m Z \, \gamma^m \, G \, \eta_+  
&=& - 16 \sqrt{2}\, g^{-1} Z^{1/4} m \frac{(z^i)^2}{r^2} \, 
\partial_l Z \, \gamma^l  \eta_-
\nonumber\\
&&- 16 \sqrt{2}\, g^{-1} Z^{1/4} m \, \frac{(z^l {\bar z}^i - z^i
{\bar z}^l)}{r^2} \, \partial_{\bar i} Z \, \gamma^{l}\, \eta_- \;.
\end{eqnarray}
With $Z = R^4/r^4$ it follows that $\partial_l Z = - 4 Z r^{-2} {\bar
z}^l$, and hence we obtain
\begin{equation}
\partial_m Z \, \gamma^m \, G \, \eta_+ = 32 \sqrt{2}\, g^{-1} Z^{5/4}
\, m \, \frac{z^l }{r^2} \, \, \gamma^{l}\, \eta_- \;.
\end{equation}
This is of the form
\begin{equation}
\partial_m Z \, \gamma^m \, G \, \eta_+ = 32 \sqrt{2} \, g^{-1}
Z^{5/4} \, R^{-1} \, m \, \chi_+ \;,
\end{equation}
where $\chi_+ = \frac12 Z^{1/4} \,w_l \, \gamma^l \, \eta_-$ and $w_l
= 2 r^{-1} z^l$.
The latter satisfies $ w_l \, {\bar w}_{\bar l} \, \delta^{l \bar l} =
2$ \cite{DallAgata:2004dk}.

\bigskip

It follows that
\begin{equation}
\epsilon_1 =\frac{\sqrt{2}}{12} \,i {\rm e}^{i \theta} \, R \, m \,
Z^{1/8} \, \varepsilon^* (x^{\mu}) \otimes \chi_+ \;,
\end{equation}
and hence
\begin{equation}
b=0 \;\;\;,\;\;\; c = \frac{\sqrt{2}}{12}  \,i {\rm e}^{i \theta}
 \, R \, m \, Z^{1/8} \;.
\end{equation}

Having identified the one-form $w$,
\begin{equation}
w = w_m\, dx^m = w_i \,dz^i = 2 \frac{z^i}{r} \, dz^i \;,
\label{wone}
\end{equation}
we proceed to decompose $P$ as in (\ref{P}).
Using (\ref{tau}) we compute
\begin{equation}
P = \frac{1}{36}  m^2 R^4 \,{\rm e}^{2 i \theta} \,
 \left(- 4 \frac{({\bar z}^l)^2}{r^6} 
\, {\bar z}^i
dz^i + \left[ 2 \frac{{\bar z}^i}{r^4} - 4 
\frac{({\bar z}^l)^2}{r^6} \, z^i \right]
d {\bar z}^i \right) \;.
\label{Ptau}
\end{equation}
Equating (\ref{Ptau}) with (\ref{P}) gives
\begin{equation}
\Pi = \frac{1}{36} m^2 R^4 \,{\rm e}^{2 i \theta} \, \left(\left[- 4
\frac{({\bar z}^l)^2}{r^6} \, {\bar z}^i - 2 {\tilde p}_1
\frac{z^i}{r} \right] dz^i + \left[ 2 \frac{{\bar z}^i}{r^4} -2
{\tilde p}_2 \frac{{\bar z}^i}{r} - 4 \frac{({\bar z}^l)^2}{r^6} \,
z^i \right] d {\bar z}^i \right) ,
\label{pip1p2}
\end{equation}
where $p_{1,2} = \frac{1}{36}  m^2 R^4 \,{\rm e}^{2 i \theta} \,
{\tilde p}_{1,2}$.
Demanding that $w \lrcorner \Pi =0$ as well as ${\bar w} \lrcorner \Pi
=0$ yields
\begin{eqnarray}
{\tilde p}_1 &=& - \frac{4}{r^3} \, \frac{({\bar z}^l)^2}{r^2} \,
\frac{({\bar z}^j)^2}{r^2} = \frac{4}{r^3} \, \left(\Sigma + i
\frac{Y}{2}\right)^2 \;, \nonumber\\
{\tilde p}_2 &=& \frac{1}{r^3} \left(1 - 4\, \frac{({\bar
z}^l)^2}{r^2} \, \frac{({ z}^j)^2}{r^2} \right) \;,
\label{p1p2}
\end{eqnarray}
where we used $dz^i \lrcorner dz^j = 0, d{\bar z}^i \lrcorner d{\bar
z}^j =0$ as well as $dz^i \lrcorner d {\bar z}^j = \delta^{i \bar j}$.
Inserting (\ref{p1p2}) back into (\ref{pip1p2}) yields
\begin{equation}
\Pi= -\frac{1}{9} m^2 R^4 \,{\rm e}^{2 i \theta} \, \frac{\left({\bar
z}^l\right)^2}{r^6} \, \left( \left({\bar z}^i - 2 \frac{({\bar
z}^j)^2}{r^2} \, z^i \right)\, dz^i + \left(z^i - 2
\frac{({z}^j)^2}{r^2} \, {\bar z}^i \right)\, d{\bar z}^i \right) \;.
\end{equation}

Now we are in position to check (\ref{gsug}). 
Using (\ref{g30u}), the lhs of (\ref{gsug}) gives
\begin{equation}
2 \sqrt{2}\,\kappa \, g^{SU}_{3,0} = - \frac{i}{27} \, {\rm e}^{i \theta} \, m^3\,
Z^2 \, r^{2} \, \left(\Sigma + i \frac{Y}{2}\right)^2 \;.
\label{lhs}
\end{equation}
The rhs  of (\ref{gsug}) yields
\begin{equation}
2 \sqrt{2}\, Z^{1/2}\, \frac{{\bar c}}{a} \, p_1 = - \frac{i}{27}\, \,
{\rm e}^{i \theta} \, m^3 \, Z^2 \,\,r^{2} \, \left(\Sigma + i
\frac{Y}{2}\right)^2 \;,
\label{rhs}
\end{equation}
Comparing (\ref{lhs}) with (\ref{rhs}) we find a perfect agreement.

\bigskip

The $(3,0)$-part of $G_3$ is thus due to the term $p_1 w$ in (\ref{P}).
It is instructive to make the this manifest in (\ref{g30ddil}).
Using
\begin{equation}
g \, d\tau|_{(1,0)} = - 2 i {\rm e}^{-2 i \theta} \left(p_1 \,w + \Pi_i \,dz^i
\right) 
\end{equation}
and ${\bar S}_2^{(2,0)} \wedge \Pi_i \,dz^i =0$, we obtain
\begin{equation}
G^{(3,0)} =  - \frac{1}{3}\,
{\rm e}^{-2 i \theta} \, \,g^{-1}\, Z \, p_1\,
{\bar S}_2^{(2,0)} \wedge w =
- \frac{1}{3} \,{\rm e}^{-2 i \theta}
\,g^{-1}\, Z \, m \,p_1\,
r \, dz^1 \wedge dz^2 \wedge dz^3 \;.
\end{equation}

Another of the supersymmetry conditions derived from the dilatino variation 
states that \cite{DallAgata:2004dk}  
\begin{equation}
p_{2}= \kappa\, Z^{-1/2}\, \frac{c}{\bar a}\; g_{21}^{SU},
\label{eq:eqgian}
\end{equation}
where 
\begin{equation}
g_{21}^{SU} = \frac{1}{16} Z^{3/2}\,  X_{mnp} \, G^{mnp}_{SU}
 \;\;\;,\;\;\;
X_{mnp} = {\overline K}_{[mn} \, w_{p]} \;,
\label{susyp2}
\end{equation}
where as before $\kappa \, G_{SU} = i g \, {\rm e}^{i \theta} \, G$.
Here $K$ is a 2--form satisfying $K \wedge w = \Omega = \frac13
\sqrt{2} \varepsilon_{ijk} dz^i \wedge dz^j \wedge dz^k$ as well as
$K_{ij} {\bar K}_{{\bar i} {\bar j}} = 8$ and also $w \lrcorner K =
{\bar w} \lrcorner K = 0$ \cite{DallAgata:2004dk}.  
Using (\ref{wone}) we therefore establish that
\begin{equation}
K = \sqrt{2}\, \varepsilon_{ij{\bar k}} \frac{{\bar z}^k}{r} dz^i
\wedge dz^j \;.
\end{equation}
It follows that
\begin{equation}
{\overline K} \wedge w = 2 \sqrt{2} \, \varepsilon_{{\bar i} {\bar j} {k}} \, 
\frac{{z}^k z^l}{r^2} d{\bar z}^i \wedge d{\bar z}^j \wedge dz^l \;,
\end{equation}
which is indeed primitive \cite{DallAgata:2004dk}, i.e. $J \wedge
{\overline K} \wedge w =0$.
The non-vanishing components of $X_{mnp}$ are therefore $X_{{\bar i}
{\bar j} l}$.  
Hence $X_{mnp} \, G^{mnp} = X_{{\bar i} {\bar j} l} \, G^{{\bar i}
{\bar j} l} = Z^{-3/2} \, X_{{\bar i} {\bar j} l} \, G_{ij \bar l}$.
Using the expressions for $G_{ij \bar l}$ given in (\ref{vv}) yields
\begin{equation}
X_{{\bar i} {\bar j} l} \, G_{ij \bar l} = \frac{8 \sqrt{2}}{3} \,
g^{-1} \, m\, Z \, r^3 \, {\tilde p}_2 \;.
\end{equation}
Now we are in position to check (\ref{eq:eqgian}). 
The rhs of (\ref{eq:eqgian}) yields
\begin{equation}
\frac{\kappa}{16} \, \frac{c}{\bar a} \, Z\, X_{mnp} \, G^{mnp}_{SU}
= \frac{1}{36} \, \, {\rm e}^{2 i \theta} \, \, m^2\, R^4 \, \,
{\tilde p}_2 = p_2 \;,
\end{equation}
so that the supersymmetry condition is indeed satisfied.
And finally, we observe that equations (3.6) and (3.7) 
given in \cite{DallAgata:2004dk} and stemming from the gaugino variation 
are also satisfied.

\bigskip

Hence we conclude that to order $m^3$ the bulk effects we computed are 
consistent with supersymmetry based on the type D ansatz (\ref{typeD}).

\section{Conclusions}

In this paper we computed order $m^3$ modifications of the
Polchinski-Strassler solution due to bulk effects.  We showed, in particular,
that a $(3,0)$--piece in $G_3$ gets generated at order $m^3$.
We argued that the associated 2--form potential may contain a term
with the polarisation tensor $\varepsilon_{ijk}$
which, in the confining phase of the dual ${\cal N}=1^*$ gauge theory,
is associated to the formation of a gaugino condensate, thereby linking
the emergence of a $G^{(3,0)}$--piece to the formation of a gaugino
condensate.  We also showed that the this $G^{(3,0)}$--piece, computed
from the bulk equation of motion for $G_3$, is consistent with the 
type D spinor ansatz introduced in \cite{DallAgata:2004dk}.  
The latter is based on the existence of a globally defined complex
vector $w$, whose asymptotic form we determined.
The existence of $w$ implies that the transverse
6--dimensional space possesses an $SU(2)$--structure \cite{DallAgata:2004dk}.
Thus, the Polchinski-Strassler solution is a concrete example of
a background with such a structure.

\bigskip

The results provided here are a further step 
towards the complete supergravity solution dual to the ${\cal N} =
1^{*}$ gauge theory.
Understanding that the full solution
possesses an $SU(2)$ structure should be helpful in simplifying the
metric and flux ansaetze needed in order to obtain the full solution.
It should also be noted that a candidate for the complete metric is
given in \cite{Pilch:2000fu}, where the authors uplifted the
5--dimensional flow solution of GPPZ \cite{Girardello:1999bd}.
Although they give the metric, they did not determine $G_{3}$.
We hope that our results will be useful in achieving this.
Since \cite{Pilch:2000fu} is the uplift of the 5--dimensional GPPZ
solution, it can describe the same physics as \cite{Polchinski:2000uf} 
if all the massive KK modes vanish along the flow.
As discussed in section 4, some of these modes may however be turned
on, since we have encountered them as homogeneous solutions of the
equations of motion for $G_{3}$, with overall coefficients which are
only fixed by the infrared physics.

\bigskip

\bigskip \bigskip
\noindent
{\bf Acknowledgements}

\bigskip

We would like to thank P. Di Vecchia, J. Erdmenger, Z. Guralnik, K.
Pilch, N. Prezas and M. Strassler for valuable comments and discussions.
The work of G.L.C.\ is supported by the DFG.
Work supported in part by the European Community's Human Potential
Programme under contract HPRN-CT-2000-00131 Quantum Spacetime.

\appendix

\section{$T_3$ and $V_3$ \label{g3m}}

Here we review some of the properties of $G_3 = d \eta_2 = \zeta
 \, g^{-1} \,
Z \, \left(T_3 - \frac43 V_3 \right)$ \cite{Polchinski:2000uf}.  We will set $\zeta =1$
in the following.  It may be restored by rescaling $g^{-1}$.

\bigskip

$T_3$ is given by
(\ref{t3}).
It has Hodge type $(1,2)$ and is primitive, 
i.e. $T_3 \wedge J = 0$.  
It satisfies $3 T_3 = d S_2$ with \begin{eqnarray} S_2  &= & m \,
\left( z^1  d {\bar z}^2 \wedge d {\bar z}^3 + {\bar z}^1  d z^2
\wedge d {\bar z}^3 +
  {\bar z}^1   d {\bar z}^2 \wedge d z^3 \right. \nonumber\\
&& \qquad -  \left. z^2  d {\bar z}^1 \wedge d {\bar z}^3 - {\bar
z}^2  d z^1 \wedge d {\bar z}^3 -
  {\bar z}^2   d {\bar z}^1 \wedge d z^3 \right. \nonumber\\
&& \qquad + \left. z^3  d {\bar z}^1 \wedge d {\bar z}^2 + {\bar
z}^3  d z^1 \wedge d {\bar z}^2 +
  {\bar z}^3   d {\bar z}^1 \wedge d z^2
 \right) \;.
\label{s2}
\end{eqnarray}
$V_3$ is given by 
$V_3 = d \log r \wedge S_2= (2 r^2)^{-1} d r^2 \wedge
S_2$.  $G_3$ can then be written as
$G_3 = (3 g)^{-1}R^4 \, d \left(r^{-4} S_2
\right)$.

\bigskip

$T_3$ and $V_3$ satisfy the following relations,
\begin{eqnarray}
\star_6 \,T_3 &=& - i T_3 \;,\nonumber\\
{\star_6 }\, V_3 &=& - i (T_3 - V_3 ) \;, \nonumber\\
(\star_6 - i )\, V_3 &=& - i T_3 \;, \nonumber\\
(\star_6 + i )\, V_3 &=& - i (T_3- 2V_3) \;, \nonumber\\
(\star_6 - i )\, d \eta_2 &=& -\frac{2i}{3} \,
g^{-1} \, Z \, T_3 \;, \nonumber\\
(\star_6 + i )\, d \eta_2 &=&  \frac{4i}{3} \, g^{-1} \, Z \,
\left(T_3 - 2 V_3 \right)
\;, \nonumber\\
(\star_6 + i ) d {\bar \eta}_2 &=& \frac{2i}{3} \, g^{-1} \, Z \,
{\bar T}_3
\;, \nonumber\\
(\star_6 + i )\, {\rm Im } \, d \eta_2 &=& \frac{2}{3} \, g^{-1} \, Z
\, \left(T_3 - 2 V_3 \right) - \frac{1}{3} \, g^{-1} \, Z \,
{\bar T}_3
\;, \nonumber\\
{\star_6 }\, d \eta_2 &=& \frac{i}{3}\, g^{-1} \, Z \, \left( T_3 - 4
V_3 \right)
\;, \nonumber\\
(\star_6 + i )\, {\rm Im } \,d \eta_2 + i \star_6 d \eta_2 &=& \frac13\,
g^{-1} \, Z \, (T_3 - {\bar T}_3) \;. \label{rela} 
\end{eqnarray}  
Inspection of (\ref{hodge}) and of (\ref{rela}) shows that 
$V_3$ has Hodge types $(1,2), (2,1)$ and $(0,3)$, i.e. $V_3 =
V^{(1,2)} + V^{(2,1)} + V^{(0,3)}$.  
 Using that $r^2 = 2 z^i {\bar
  z}^i$ and $J = i (dz^1 \wedge d {\bar z}^1 +
dz^2 \wedge d {\bar z}^2+ dz^3 \wedge d {\bar z}^3)  $,
we compute $V_3 = d \log r \wedge S_2= (2 r^2)^{-1} d r^2 \wedge
S_2$ and obtain \begin{eqnarray}
V^{(1,2)} &=& \frac12 \, T_3 + \Delta^{(1,2)}_{NP} \;, \nonumber\\
\Delta^{(1,2)}_{NP} &=& -i \,\frac{ m}{r^2} \,
\varepsilon_{i{\bar j} {\bar k}}\, z^i\, {\bar z}^j \,d {\bar z}^k
\wedge J \;,\nonumber\\
 V_P^{(2,1)} &=& \frac{m}{ r^2} \, \left( \; \;\; [({\bar z}^1)^2 +
({\bar z}^2)^2] \, dz^1 \wedge dz^2 \wedge d {\bar z}^3 
\right. \nonumber\\
&& \left. \qquad + [({\bar z}^1)^2 +
({\bar z}^3)^2] \, dz^1 \wedge d {\bar z}^2 \wedge d {z}^3 \right. \nonumber\\
&&\left. \qquad +  [({\bar z}^2)^2 + ({\bar z}^3)^2] \,
d {\bar z}^1 \wedge dz^2 \wedge dz^3 
\right. \nonumber\\
&& \left. 
\qquad - {\bar z}^1 {\bar z}^3 \, dz^2 \wedge ( d {z}^3
\wedge d {\bar z}^3 - dz^1 \wedge d {\bar
  z}^1 ) \right. \nonumber\\
&& \left. \qquad +{\bar z}^1 {\bar z}^2 \, dz^3 \wedge (d {z}^2
\wedge d {\bar z}^2 - dz^1  \wedge d {\bar
  z}^1 ) 
\right. \nonumber\\
&& \left. 
\qquad + {\bar z}^2 {\bar z}^3 \,dz^1 \wedge (d {z}^3
\wedge d {\bar z}^3 - dz^2 \wedge d {\bar
  z}^2 )
\right)
\;, \nonumber\\
V^{(0,3)} &=& m \, \frac{(z^i)^2}{r^2} \, d {\bar z}^1 \wedge d
{\bar z}^2 \wedge d {\bar z}^3 \;. \label{vv} 
\end{eqnarray} 
We also note the following useful relations,
\begin{eqnarray} 
(\star_6 - i ) \, \Delta^{(1,2)}_{NP} &=& 0 \;,
\nonumber\\
(\star_6 - i ) \, V^{(1,2)} &=& -i T_3 \;. 
\end{eqnarray} 
The expressions (\ref{vv})
are in agreement with the findings of \cite{Grana:2000jj}
based on supersymmetry.
Namely, setting $W = m \, (z^i)^2$, we have \begin{eqnarray} Z \,
\Delta^{(1,2)}_{NP} &\propto& \varepsilon_{ijk}\, A_{ij}\,d {\bar
z}^k \;\;\;,\;\;\;   A_{ij} =
\partial_{[i} W \, \partial_{j]} Z \;, \nonumber\\
Z \, V^{(0,3)} &\propto& \partial_i W \, \partial_{\bar i} Z \, d
{\bar z}^1 \wedge d {\bar z}^2 \wedge d {\bar z}^3 \;, \nonumber\\
Z \, V_P^{(2,1)} &\propto& S_{{\bar i}{ \bar j}} \;\;\;,\;\;\;
S_{{\bar 1}{ \bar 1}} =  \frac{m Z}{r^2} [({\bar z}^2)^2 + ({\bar
z}^3)^2] \;\;\;,\;\;\; S_{{\bar 1}{ \bar 2}} =  \frac{m Z}{r^2}
{\bar z}^1\,{\bar z}^2 \;.
\label{grapol}
\end{eqnarray}


\begingroup\begin{thebibliography}{10}

\bibitem{Maldacena:1998re}
J.~M. Maldacena, {\it The large N limit of superconformal field theories and
  supergravity},  {\em Adv. Theor. Math. Phys.} {\bf 2} (1998) 231--252
  [\href{http://arXiv.org/abs/hep-th/9711200}{{\tt hep-th/9711200}}].

\bibitem{Gubser:1998bc}
S.~S. Gubser, I.~R. Klebanov and A.~M. Polyakov, {\it Gauge theory correlators
  from non-critical string theory},  {\em Phys. Lett.} {\bf B428} (1998)
  105--114 [\href{http://arXiv.org/abs/hep-th/9802109}{{\tt hep-th/9802109}}].

\bibitem{Witten:1998qj}
E.~Witten, {\it Anti-de Sitter space and holography},  {\em Adv. Theor. Math.
  Phys.} {\bf 2} (1998) 253--291
  [\href{http://arXiv.org/abs/hep-th/9802150}{{\tt hep-th/9802150}}].

\bibitem{Polchinski:2000uf}
J.~Polchinski and M.~J. Strassler, {\it The string dual of a confining
  four-dimensional gauge theory},
  \href{http://arXiv.org/abs/hep-th/0003136}{{\tt hep-th/0003136}}.

\bibitem{Vafa:1994tf}
C.~Vafa and E.~Witten, {\it A strong coupling test of S duality},  {\em Nucl.
  Phys.} {\bf B431} (1994) 3--77
  [\href{http://arXiv.org/abs/hep-th/9408074}{{\tt hep-th/9408074}}].

\bibitem{Donagi:1996cf}
R.~Donagi and E.~Witten, {\it Supersymmetric Yang-Mills theory and integrable
  systems},  {\em Nucl. Phys.} {\bf B460} (1996) 299--334
  [\href{http://arXiv.org/abs/hep-th/9510101}{{\tt hep-th/9510101}}].

\bibitem{Strassler:1998ny}
M.~J. Strassler, {\it Messages for QCD from the superworld},  {\em Prog. Theor.
  Phys. Suppl.} {\bf 131} (1998) 439--458
  [\href{http://arXiv.org/abs/hep-lat/9803009}{{\tt hep-lat/9803009}}].

\bibitem{Dorey:1999sj}
N.~Dorey, {\it An elliptic superpotential for softly broken N = 4
  supersymmetric Yang-Mills theory},  {\em JHEP} {\bf 07} (1999) 021
  [\href{http://arXiv.org/abs/hep-th/9906011}{{\tt hep-th/9906011}}].

\bibitem{Dorey:2000fc}
N.~Dorey and S.~P. Kumar, {\it Softly-broken N = 4 supersymmetry in the
large-N
  limit},  {\em JHEP} {\bf 02} (2000) 006
  [\href{http://arXiv.org/abs/hep-th/0001103}{{\tt hep-th/0001103}}].

\bibitem{Giddings:2001yu}
S.~B. Giddings, S.~Kachru and J.~Polchinski, {\it Hierarchies from fluxes in
  string compactifications},  {\em Phys. Rev.} {\bf D66} (2002) 106006
  [\href{http://arXiv.org/abs/hep-th/0105097}{{\tt hep-th/0105097}}].

\bibitem{Ferrara:1983qs}
S.~Ferrara, L.~Girardello and H.~P. Nilles, {\it Breakdown of local
  supersymmetry through gauge fermion condensates},  {\em Phys. Lett.} {\bf
  B125} (1983) 457.

\bibitem{Dine:1985rz}
M.~Dine, R.~Rohm, N.~Seiberg and E.~Witten, {\it Gluino condensation in
  superstring models},  {\em Phys. Lett.} {\bf B156} (1985) 55.

\bibitem{Derendinger:1985kk}
J.~P. Derendinger, L.~E. Iba\~nez and H.~P. Nilles, {\it On the low-energy 
D = 4, N=1 supergravity theory extracted from the D = 10, N=1 superstring},  {\em
  Phys. Lett.} {\bf B155} (1985) 65.

\bibitem{Derendinger:1986cv}
J.~P. Derendinger, L.~E. Iba\~nez and H.~P. Nilles, {\it On the low-energy limit
  of superstring theories},  {\em Nucl. Phys.} {\bf B267} (1986) 365.

\bibitem{Kounnas:1988ye}
C.~Kounnas and M.~Porrati, {\it Spontaneous supersymmetry breaking in string
  theory},  {\em Nucl. Phys.} {\bf B310} (1988) 355.

\bibitem{Font:1990nt}
A.~Font, L.~E. Iba\~nez, D.~L\"ust and F.~Quevedo, {\it Supersymmetry breaking from
  duality invariant gaugino condensation},  {\em Phys. Lett.} {\bf B245} (1990)
  401--408.

\bibitem{Ferrara:1990ei}
S.~Ferrara, N.~Magnoli, T.~R. Taylor and G.~Veneziano, {\it Duality and
  supersymmetry breaking in string theory},  {\em Phys. Lett.} {\bf B245}
  (1990) 409--416.

\bibitem{Nilles:1990jv}
H.~P. Nilles and M.~Olechowski, {\it Gaugino condensation and duality
  invariance},  {\em Phys. Lett.} {\bf B248} (1990) 268--272.

\bibitem{Cardoso:2003sp}
G.~L. Cardoso, G.~Curio, G.~Dall'Agata and D.~L\"ust, {\it Heterotic string
  theory on {non-Kaehler} manifolds with {H}- flux and gaugino condensate},
  {\em Fortsch. Phys.} {\bf 52} (2004) 483--488
  [\href{http://arXiv.org/abs/hep-th/0310021}{{\tt hep-th/0310021}}].

\bibitem{Strominger:1986uh}
A.~Strominger, {\it Superstrings with torsion},  {\em Nucl. Phys.} {\bf B274}
  (1986) 253.

\bibitem{Cardoso:2002hd}
G.~L. Cardoso, G.~Dall'Agata, D.~L\"ust, P.~Manousselis and G.~Zoupanos, {\it
  {Non-Kaehler} string backgrounds and their five torsion classes},  {\em Nucl.
  Phys.} {\bf B652} (2003) 5--34
  [\href{http://arXiv.org/abs/hep-th/0211118}{{\tt hep-th/0211118}}].

\bibitem{Becker:2003yv}
K.~Becker, M.~Becker, K.~Dasgupta and P.~S. Green, {\it Compactifications of
  heterotic theory on {non-Kaehler} complex manifolds. {I}},  {\em JHEP} {\bf
  04} (2003), no.~007 052 [\href{http://arXiv.org/abs/hep-th/0301161}{{\tt
  hep-th/0301161}}].

\bibitem{Becker:2003gq}
K.~Becker, M.~Becker, K.~Dasgupta and S.~Prokushkin, {\it Properties of
  heterotic vacua from superpotentials},  {\em Nucl. Phys.} {\bf B666} (2003)
  144--174 [\href{http://arXiv.org/abs/hep-th/0304001}{{\tt hep-th/0304001}}].

\bibitem{Cardoso:2003af}
G.~L. Cardoso, G.~Curio, G.~Dall'Agata and D.~L\"ust, {\it {BPS} action and
  superpotential for heterotic string compactifications with fluxes},  {\em
  JHEP} {\bf 10} (2003) 004 [\href{http://arXiv.org/abs/hep-th/0306088}{{\tt
  hep-th/0306088}}].

\bibitem{Becker:2003sh}
K.~Becker, M.~Becker, P.~S. Green, K.~Dasgupta and E.~Sharpe, {\it
  Compactifications of heterotic strings on {non-Kaehler} complex manifolds.
  {II}},  {\em Nucl. Phys.} {\bf B678} (2004) 19--100
  [\href{http://arXiv.org/abs/hep-th/0310058}{{\tt hep-th/0310058}}].

\bibitem{Gauntlett:2003cy}
J.~P. Gauntlett, D.~Martelli and D.~Waldram, {\it Superstrings with intrinsic
  torsion},  \href{http://arXiv.org/abs/hep-th/0302158}{{\tt hep-th/0302158}}.

\bibitem{Camara:2003ku}
P.~G. Camara, L.~E. Iba\~nez and A.~M. Uranga, {\it Flux-induced {SUSY-breaking}
  soft terms},  {\em Nucl. Phys.} {\bf B689} (2004) 195--242
  [\href{http://arXiv.org/abs/hep-th/0311241}{{\tt hep-th/0311241}}].

\bibitem{Grana:2003ek}
M.~Gra\~na, T.~W. Grimm, H.~Jockers and J.~Louis, {\it Soft supersymmetry
  breaking in Calabi-Yau orientifolds with D-branes and fluxes},
  \href{http://arXiv.org/abs/hep-th/0312232}{{\tt hep-th/0312232}}.

\bibitem{Freedman:2000xb}
D.~Z. Freedman and J.~A. Minahan, {\it Finite temperature effects in the
  supergravity dual of the $N = 1^{*}$ gauge theory},  {\em JHEP} {\bf 01} (2001)
  036 [\href{http://arXiv.org/abs/hep-th/0007250}{{\tt hep-th/0007250}}].

\bibitem{Becker:1996gj}
K.~Becker and M.~Becker, {\it {M-Theory} on eight-manifolds},  {\em Nucl.
  Phys.} {\bf B477} (1996) 155--167
  [\href{http://arXiv.org/abs/hep-th/9605053}{{\tt hep-th/9605053}}].

\bibitem{Grana:2000jj}
M.~Gra\~na and J.~Polchinski, {\it Supersymmetric three-form flux perturbations
  on {$AdS_5$}},  {\em Phys. Rev.} {\bf D63} (2001) 026001
  [\href{http://arXiv.org/abs/hep-th/0009211}{{\tt hep-th/0009211}}].

\bibitem{Grana:2001xn}
M.~Gra\~na and J.~Polchinski, {\it Gauge / gravity duals with holomorphic
  dilaton},  {\em Phys. Rev.} {\bf D65} (2002) 126005
  [\href{http://arXiv.org/abs/hep-th/0106014}{{\tt hep-th/0106014}}].

\bibitem{Frey:2003sd}
A.~R. Frey and M.~Gra\~na, {\it Type {IIB} solutions with interpolating
  supersymmetries},  {\em Phys. Rev.} {\bf D68} (2003) 106002
  [\href{http://arXiv.org/abs/hep-th/0307142}{{\tt hep-th/0307142}}].

\bibitem{DallAgata:2004dk}
G.~Dall'Agata, {\it On supersymmetric solutions of type IIB supergravity with
  general fluxes},  \href{http://arXiv.org/abs/hep-th/0403220}{{\tt
  hep-th/0403220}}.

\bibitem{Frey:2004rn}
A.~R. Frey, {\it Notes on SU(3) structures in type IIB supergravity},
  \href{http://arXiv.org/abs/hep-th/0404107}{{\tt hep-th/0404107}}.

\bibitem{DallAgata:2003ir}
G.~Dall'Agata and N.~Prezas, {\it {N = 1} geometries for {M-theory} and type
  {IIA} strings with fluxes},  {\em Phys. Rev.} {\bf D69} (2004) 066004
  [\href{http://arXiv.org/abs/hep-th/0311146}{{\tt hep-th/0311146}}].

\bibitem{Gauntlett:2004zh}
J.~P. Gauntlett, D.~Martelli, J.~Sparks and D.~Waldram, {\it Supersymmetric
  {$AdS_5$} solutions of {M-theory}},
  \href{http://arXiv.org/abs/hep-th/0402153}{{\tt hep-th/0402153}}.

\bibitem{Myers:1999ps}
R.~C. Myers, {\it Dielectric-branes},  {\em JHEP} {\bf 12} (1999) 022
  [\href{http://arXiv.org/abs/hep-th/9910053}{{\tt hep-th/9910053}}].

\bibitem{Pilch:2004yg}
K.~Pilch and N.~P. Warner, {\it {N=1} supersymmetric solutions of {IIB}
  supergravity},  \href{http://arXiv.org/abs/hep-th/0403005}{{\tt
  hep-th/0403005}}.

\bibitem{Kim:1985ez}
H.~J. Kim, L.~J. Romans and P.~van Nieuwenhuizen, {\it The mass spectrum of
  chiral {N=2 D = 10} supergravity on $S^5$},  {\em Phys. Rev.} {\bf D32}
  (1985) 389.

\bibitem{Ceresole:1999zs}
A.~Ceresole, G.~Dall'Agata, R.~D'Auria and S.~Ferrara, {\it Spectrum of type
  {IIB} supergravity on {$AdS_5 \times T^{11}$}: Predictions on {N = 1}
  {SCFT's}},  {\em Phys. Rev.} {\bf D61} (2000) 066001
  [\href{http://arXiv.org/abs/hep-th/9905226}{{\tt hep-th/9905226}}].

\bibitem{Pilch:2000fu}
K.~Pilch and N.~P. Warner, {\it N = 1 supersymmetric renormalization group
  flows from IIB supergravity},  {\em Adv. Theor. Math. Phys.} {\bf 4} (2002)
  627--677 [\href{http://arXiv.org/abs/hep-th/0006066}{{\tt hep-th/0006066}}].

\bibitem{Girardello:1999bd}
L.~Girardello, M.~Petrini, M.~Porrati and A.~Zaffaroni, {\it The supergravity
  dual of N = 1 super Yang-Mills theory},  {\em Nucl. Phys.} {\bf B569} (2000)
  451--469 [\href{http://arXiv.org/abs/hep-th/9909047}{{\tt hep-th/9909047}}].

\end{thebibliography}\endgroup

\providecommand{\href}[2]{#2}

\end{document}